\documentclass[
aps,prx,reprint,superscriptaddress
]{revtex4-2}

\usepackage{amsmath}
\usepackage{mathtools}
\usepackage{comment}
\usepackage{graphicx}
\usepackage{dcolumn}
\usepackage{bm}
\usepackage{xspace}
\usepackage{physics}
\usepackage{siunitx}
\usepackage[dvipsnames]{xcolor}
\usepackage[version=3]{mhchem}

\usepackage{newfloat}

\DeclareFloatingEnvironment[name={Supplementary Figure},fileext=lof]{suppfigure}
\DeclareFloatingEnvironment[name={Supplementary Table},fileext=lof]{supptable}

\DeclareSIUnit\angstrom{\text{Å}}
\DeclareMathOperator{\sgn}{sgn}
\newcommand{\RCX}{$RE_8\mathrm{Co}X_3$\xspace}
\newcommand{\YCI}{$\mathrm{Y}_8\mathrm{CoIn}_3$\xspace}
\newcommand{\boldrm}[1]{\bm{\mathrm{#1}}}

\usepackage{txfonts}
\usepackage{amsfonts}
\usepackage{color}%
\usepackage[normalem]{ulem}

\definecolor{myorange}{rgb}{0.7,0.5,0.0}
\definecolor{mygreen}{rgb}{0.0,0.7,0.0}
\definecolor{purple}{rgb}{0.75,0.0,1.0}

\newcommand{\tcr}[1]{\textcolor{black}{#1}}

\newcommand{\AppliedPhys}{Department of Applied Physics, The University of Tokyo, Bunkyo, Tokyo 113-8656, Japan}
\newcommand{\RIKEN}{RIKEN Center for Emergent Matter Science, Wako, Saitama 351-0198, Japan}
\newcommand{\RIKENCPR}{Condensed Matter Theory Laboratory, RIKEN CPR, Wako, Saitama 351-0198, Japan}
\newcommand{\BasicScience}{Department of Basic Science, The University of Tokyo, Meguro, Tokyo 153-8902, Japan}
\newcommand{\Komaba}{Komaba Institute for Science, The University of Tokyo, Meguro, Tokyo 153-8902, Japan}
\newcommand{\Julich}{Peter Gr\"{u}nberg Institut and Institute for Advanced Simulation, Forschungszentrum J\"{u}lich and JARA, D-52425 Jülich, Germany}

\begin{document}

\title{
Ideal Spin-Orbit-Free Dirac Semimetal and Diverse Topological Transitions in Pr$_8$CoGa$_3$ Family
}

\author{Manabu Sato} \affiliation{\AppliedPhys} 
\author{Juba Bouaziz}  \affiliation{\Julich}
\author{Shuntaro Sumita} \affiliation{\BasicScience} \affiliation{\Komaba} \affiliation{\RIKENCPR}
\author{Shingo Kobayashi} \affiliation{\RIKEN}
\author{Ikuma Tateishi} \affiliation{\RIKEN}
\author{Stefan Bl\"{u}gel} \affiliation{\Julich}
\author{Akira Furusaki} \affiliation{\RIKENCPR} \affiliation{\RIKEN}
\author{Motoaki Hirayama}       \email{hirayama@ap.t.u-tokyo.ac.jp}   \affiliation{\AppliedPhys} \affiliation{\RIKEN}

\date{\today}

\begin{abstract}
Topological semimetals, known for their intriguing properties arising from band degeneracies, have garnered significant attention.
However, the discovery of a material realization and the detailed characterization of spinless Dirac semimetals have not yet been accomplished.
Here, we propose from first-principles calculations that the $RE_8\mathrm{Co}X_3$ group ($RE$ = rare earth elements, $X$ = Al, Ga, or In) contains ideal spinless Dirac semimetals whose Fermi surfaces are fourfold degenerate band-crossing points (without including spin degeneracy).
Despite the lack of space inversion symmetry in these materials, Dirac points are formed on the rotation-symmetry axis due to accidental degeneracies of two bands corresponding to different 2-dimensional irreducible representations of $C_{6v}$ group.
We also investigate, through first-principles calculations and effective model analysis, various phase transitions caused by lattice distortion or elemental substitutions from the Dirac semimetal phase to distinct topological semimetallic phases such as nonmagnetic linked-nodal-line and Weyl semimetals (characterized by the second Stiefel-Whitney class) and ferromagnetic Weyl semimetals.
\end{abstract}

\maketitle

\section{INTRODUCTION}

Topological semimetals are a fascinating group of materials with degeneracies between valence and conduction bands that exhibit intriguing properties in the bulk and at the surface, e.g. due to interference effects between the electronic states around the degeneracy points~\cite{RevModPhys.90.015001}.
There exist various types of topological semimetals in real materials including Weyl~\cite{Xu2015,Yang2015,PhysRevX.5.031013,PhysRevX.5.011029,PhysRevLett.114.206401}, Dirac~\cite{PhysRevLett.108.140405,Na3Bi_exp,PhysRevB.85.195320,Neupane2014,Liu2014,Jeon2014,PhysRevLett.113.027603,PhysRevB.88.125427}, and nodal-line semimetals~\cite{PhysRevLett.115.036806,PhysRevLett.115.036807,Hirayama2017,PhysRevB.96.155206,Zhou_2020,PhysRevB.85.115105,PhysRevB.86.085149,Chen2015}, for spinless and spinful systems.
\tcr{Here spinless systems refer to electron systems without spin-orbit couplings (SOC), where the spin degrees of freedom can be omitted due to the spin degeneracy, while spinful systems have the SOC, which breaks SU(2) spin rotation symmetry.}
\tcr{Although all real materials have finite SOC, it is reasonable to first discuss the topology as spinless systems and then examine the effect of the SOC when studying materials with weak SOC.
This is because there are topological invariants that take nontrivial values only in systems without the SOC, such as the quantized Berry phase protected by a combination of space inversion and time reversal symmetry~\cite{PhysRevB.92.081201,Fang_2016,PhysRevB.96.155105}.}
\tcr{Hereafter we will disregard the spin degrees of freedom when counting the degree of degeneracy in spinless systems.}

The recent development of the representation theory of electronic energy bands, such as symmetry-based indicators~\cite{PhysRevX.7.041069,Po2017} and topological quantum chemistry~\cite{Bradlyn2017}, has enabled the exhaustive search over crystal databases for topological materials~\cite{Tang2019, Zhang2019,Vergniory2019,doi:10.1126/sciadv.aau8725,PhysRevB.100.195108,doi:10.1126/science.abg9094}.
Nevertheless, the comprehensive classification and detailed characterization of topological semimetals are still lacking.
For example, no material realization of spinless Dirac semimetals with a quadruple \tcr{degenerate point} between the valence and conduction bands has been discovered to the best of our knowledge.
\tcr{The difference between the spinless Dirac semimetal and other well-known topological semimetals with quadruple degeneracy is given in Supplementary Note 1.}
In fact, the irreducible representations (irreps) of valence bands at high symmetry points alone cannot tell us whether the material in question is an ideal semimetal without superfluous Fermi surfaces of finite area, while they can determine the presence of nodes and their degree of degeneracy~\cite{Tang2019, Zhang2019,Vergniory2019,doi:10.1126/sciadv.aau8725,PhysRevB.100.195108,doi:10.1126/science.abg9094,PhysRevX.8.031069}.
Moreover, for spinless systems, there is currently no guiding principle for the realization of band inversion, which is essential for the design of topological insulators and topological semimetals, in contrast to the spinful systems, where the band inversion can be achieved by strong SOC.
Thus, the study of spinless topological materials lags behind that of spinful ones.

\tcr{In the search for material realizations of the symmetry-protected Dirac semimetals, the representation theory of space groups is essential.}
\tcr{In spinful systems}, there are two distinct classes of Dirac semimetals~\cite{Yang2014,YangMorimotoFurusaki2015,RevModPhys.90.015001}.
In one class of Dirac semimetals, a pair of Dirac points is formed due to an accidental band crossing of two different 2-dimensional irreps on a high-symmetry line, as seen in $\mathrm{Na}_3\mathrm{Bi}$~\cite{Na3Bi_exp,PhysRevB.85.195320} and $\mathrm{Cd}_3\mathrm{As}_2$~\cite{Neupane2014,Liu2014,Jeon2014,PhysRevLett.113.027603,PhysRevB.88.125427}.
In the second class of Dirac semimetals with nonsymmorphic symmetry, such as $\beta$-$\mathrm{BiO}_2$~\cite{PhysRevLett.108.140405}, a Dirac point appears as a 4-dimensional irrep at a high-symmetry point on the boundary of the Brillouin zone (BZ).
These two classes of Dirac semimetals are also expected to exist in spinless systems, and the space groups that can realize each class are listed in Ref.~\cite{YU2022375}.
\tcr{We note that the study of spinful Dirac semimetals is carried out using the double-valued representations, while the study of spinless Dirac semimetals is based on the single-valued representations.
This corresponds to the fact that a spinless Dirac point does not consist of the spin degrees of freedom, but only of the orbital (pseudospin) degrees of freedom.
In other words, in spinless Dirac semimetals, eight bands including the spin degrees of freedom form the degeneracy, which clearly distinguishes spinless Dirac semimetals from already well-studied spinful Dirac semimetals.}

In this paper we study the first class of spinless Dirac semimetals and its material realization.
We focus on the space groups with $C_{6v}$ symmetry that accommodates two 2-dimensional irreps along a high symmetry line in the BZ.
In addition to the space group considerations, we utilize the chemical bonding perspective to design ideal Dirac semimetals.
We choose elements with similar electronegativity that form covalent bonds, which we expect to result in the reduction of the density of states (DOS) at the Fermi energy.

Our study reveals that materials within the \RCX group ($RE$ = rare earth elements, $X$ = Al, Ga, or In) exhibit the spinless Dirac semimetal phase with nontrivial topological properties.
These materials have $C_{6v}$ symmetry on the high symmetry line in the BZ.
We will begin by discussing these materials' crystal structure and their symmetry.
Taking \YCI as an example, we will present the bulk electronic structure and the characterization of its topology, and describe the surface states and their correspondence to the bulk topological invariant.
Next, we will show the diverse topological phase transitions caused by symmetry reductions in the target materials, from the Dirac semimetallic phase to nodal-line and Weyl semimetallic phases characterized by the second Stiefel-Whitney (SW) class.
Finally we will consider the substitution of rare earth elements to introduce strong SOC and discuss possible magnetic properties.

\section{RESULTS}
\subsection{Electronic band structure}

\begin{figure}[htp]
\includegraphics[width=8cm]{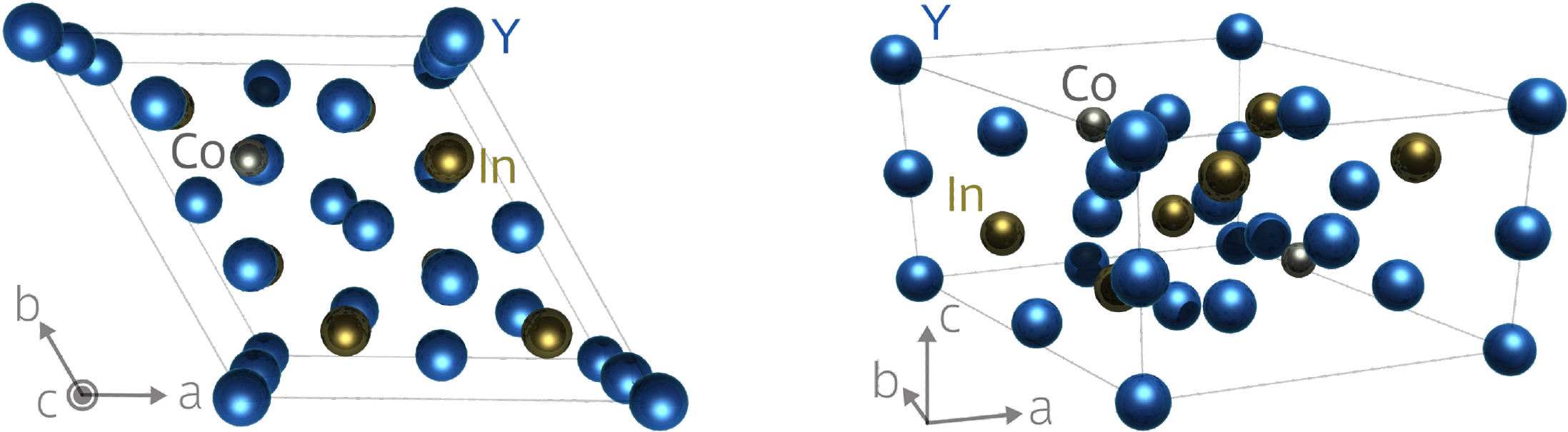}
\caption{\label{fig:Crystal}
\textbf{Crystal structure of \ce{Y8CoIn3}.} The blue, gray, and gold balls represent Y, Co, and In atoms, respectively.
}
\end{figure}

Among the ternary compounds \RCX ($RE$ = rare earth elements, $X$ = Al, Ga, or In), the experimentally synthesized compounds all show a hexagonal crystal structure with the nonsymmorphic space group $P6_3mc$ (No.~186)~\cite{DzevenkoHamykTyvanchukKalychak+2013+604+609, DZEVENKO201314,Kristallografiya}.
This structure lacks space inversion symmetry and has a polarization along the $c$-axis.
Figure~\ref{fig:Crystal} shows the crystal structure of \YCI~\cite{DZEVENKO201314}, where the measured lattice constants are $a = b = \SI{10.3678}{\angstrom}$ and $c = \SI{7.0069}{\angstrom}$.
The crystal structure is three-dimensional
and the composition is primarily made up of Y elements.
The space group has a sixfold screw symmetry around the $z$ axis $\{C_{6z}|00\frac{1}{2}\}$, a mirror symmetry with respect to the $yz$ plane $\{M_x|000\}$, and a glide symmetry with respect to the $zx$ plane $\{M_y|00\frac{1}{2}\}$, where the $z$ direction is taken to be along the $c$-axis.
As a result, the little co-group $C_{6v}$ is formed along the $k_z$ axis ($\Gamma$-A line in the BZ).

\begin{figure*}[htp]
\includegraphics[width=17cm]{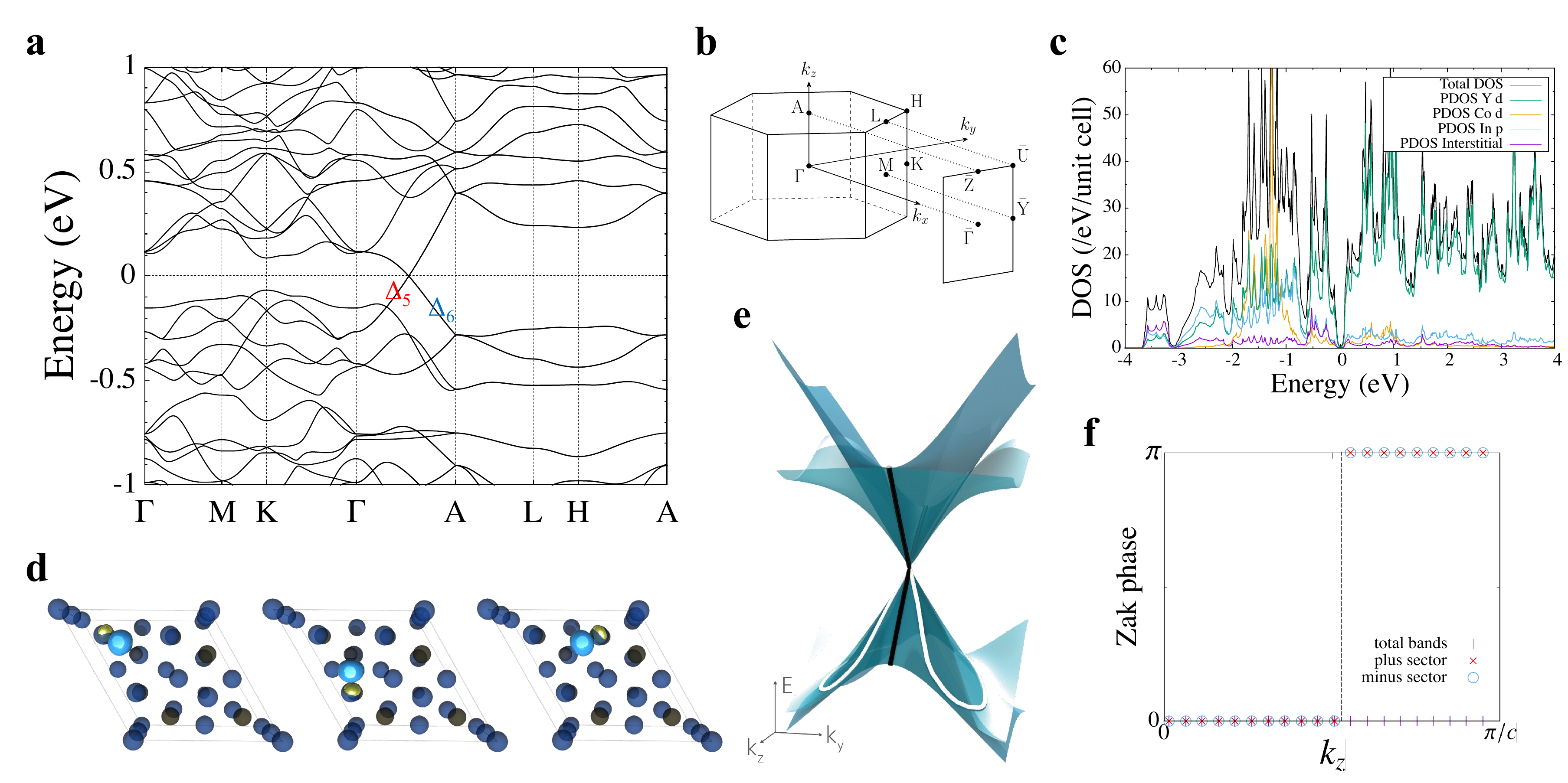}
\caption{\label{fig:SpinlessBand}
\textbf{The electronic structure of Y$_8$CoIn$_3$ calculated without the SOC.}
\textbf{a} The electronic band structure of \YCI calculated by the generalized gradient approximation (GGA).
The irreps of the two bands forming the Dirac point in the $\Gamma$-A line are shown.
The energy is measured from the Fermi level.
\textbf{b} The bulk BZ and the projected surface BZ for the (1000) surface.
\textbf{c} The total and partial density of states (PDOS) calculated with the Wannier functions.
\textbf{d} The interstitial Wannier functions near the Co atoms.
The differently colored surfaces represent isosurfaces of opposite sign.
\textbf{e} The 3-dimensional band structure around the Dirac point calculated in the $k_x=0$ plane. The black (white) lines represent degeneracies between the valence bands or between the conduction bands on (off) the $k_z$ axis.
\textbf{f} The Zak phases along the $x$ axis for each glide sector. The dashed line represents the coordinate of the Dirac point.
}
\end{figure*}

Figure~\ref{fig:SpinlessBand}a shows the electronic band structure of \YCI calculated in the absence of the SOC,
with the corresponding bulk BZ shown in Fig.~\ref{fig:SpinlessBand}b.
Clearly, \YCI is an ideal semimetal with the band crossing along the $\Gamma$-$A$ line.
The Fermi surface of this semimetal consists of two Fermi points on the $k_z$ axis that are related by time reversal symmetry.
\tcr{This band structure is well reproduced by the Korringa-Kohn-Rostoker (KKR) method (see Supplementary Note 2).}
The density of states shown in Fig.~\ref{fig:SpinlessBand}c suggests that the bands near the Fermi level originate mainly from the Y $4d$ orbitals, the Co $3d$ orbitals, and the In $5p$ orbitals.
\tcr{We note that the Y atoms occupy four Wyckoff positions, and all of them contribute almost equally to the DOS.}
Even though \YCI is composed of metallic elements, covalent bonds are formed between the Y $4d$, Co $3d$, and In $5p$ orbitals~\cite{6919084}, resulting in semimetallic density of states.
In addition to these atomic orbital states, the bands in the immediate vicinity of the Fermi level also originate from three interstitial states near the Co atoms [Fig.~\ref{fig:SpinlessBand}d], which are related to each other by the threefold rotational symmetry around the Co atoms.

Along the $\Gamma$-$A$ line, there is a little co-group $C_{6v}$. This group has two distinct 2-dimensional irreps, $\Delta_5$ and $\Delta_6$, which are distinguished by eigenvalues of the screw symmetry around the $z$ axis.
This situation is similar to a benzene molecule with $C_{6v}$, where the highest occupied molecular orbital (HOMO) and lowest unoccupied molecular orbital (LUMO) have double degeneracy of different irreps.
By contrast, $C_{3v}$ and $C_{4v}$ symmetries cannot have two distinct types of 2-dimensional irreps in spinless systems.
The two bands that intersect along the $\Gamma$-$A$ line belong to the two different 2-dimensional irreps $\Delta_5$ and $\Delta_6$ [Fig.~\ref{fig:SpinlessBand}a].
These bands cannot hybridize, and their crossing points are fourfold degenerate Dirac points (without counting the spin degeneracy).
When the SOC is turned on, the system becomes a spinful Dirac semimetal.
The SOC's magnitude at the Dirac point can be adjusted from \SI{20}{\meV} to \SI{75}{\meV} through the elemental substitutions (see Sec.~\ref{subsubsec:Spinful} and Supplementary Note 3).

Figure~\ref{fig:SpinlessBand}e shows the 3-dimensional picture of the band structure around the Dirac point in the $k_x = 0$ plane, calculated without the SOC.
We see that the band dispersion around the Dirac point is linear in all directions.
As already mentioned, both the valence and the conduction bands are doubly degenerate on the $k_z$ axis ($\Gamma$-A line), which is indicated by the black lines in the Fig.~\ref{fig:SpinlessBand}e.
These degeneracies are lifted away from the $k_z$ axis.
To see this, we construct a low-energy $4 \times 4$ $k \cdot p$ Hamiltonian up to the second order in $k$ around the Dirac point.
We find that the two-fold degenerate valence and conduction bands split into four bands, except along the nodal lines marked in white in Fig.~\ref{fig:SpinlessBand}e,
where either valence or conduction band is two-fold degenerate; we will discuss these nodal lines in more detail below.
These features are different from the band structure of spinful Dirac semimetals with spatial inversion symmetry, such as $\mathrm{Na}_3\mathrm{Bi}$, where the bands are Kramers degenerate throughout the BZ.

Next, we discuss the topological invariants that characterize the bulk states of \YCI.
The Zak phase along the $x$-axis
\begin{equation}
\theta(k_y, k_z) = - \mathrm{i} \sum_n^{\mathrm{occ.}} \oint dk_x \ev{\pdv{k_x}}{u_n(\boldrm{k})}
\end{equation}
is quantized to either 0 or $\pi$ (mod $2\pi$) due to the mirror symmetry about the $yz$ plane.
This means that the Wannier functions are located at mirror-symmetric positions.
We find that the nontrivial topology of the bulk states of \YCI is characterized by the Zak phase $\theta(0,k_z)$ calculated in the $k_y = 0$ plane.
The $k_y = 0$ plane is invariant under the glide operation with respect to the $zx$ plane, and the wavefunctions in this plane can be chosen to be eigenstates of the glide operation.
We classify the occupied bands in the $k_y = 0$ plane into two sectors according to the sign of the eigenvalues of the glide operation, and calculate the Zak phases for each sector.
As shown in Fig.~\ref{fig:SpinlessBand}f, the Zak phases change from 0 to $\pi$ at the Dirac point in both sectors.
Note that the Zak phases calculated from all occupied bands have no jump of $\pi$ at the Dirac point, since the sum of the Zak phases for each sector is 0 (mod $2\pi$) at all $k_z$.

\begin{figure}[htp]
\includegraphics[width=8.5cm]{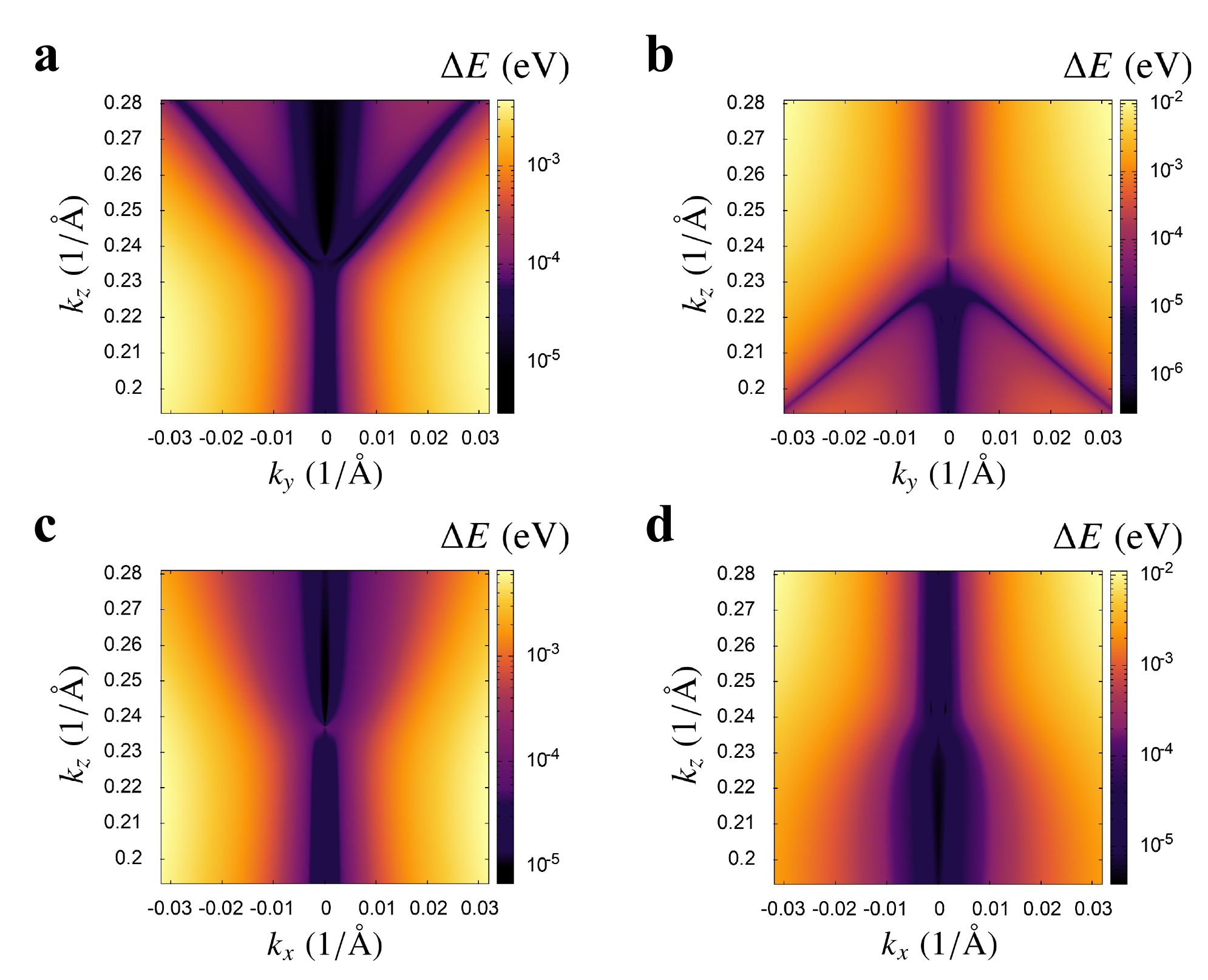}
\caption{\label{fig:degeneracy_valence}
\textbf{Color maps of the energy difference with logarithmic scale.}
The energy difference between the two topmost valence bands and between the two bottommost conduction bands of \YCI on the mirror invariant plane $k_x = 0$ are shown in \textbf{a} and \textbf{b}, respectively.
The energy difference between the two topmost valence bands and between the two bottommost conduction bands on the glide invariant plane $k_y = 0$ are shown in \textbf{c} and \textbf{d}, respectively.
}
\end{figure}

As shown in Fig.~\ref{fig:SpinlessBand}e, there are lines of degeneracies between two valence bands and between two conduction bands, which are connected to the Dirac point. These nodal lines are found not only on the $k_z$ axis but also on the mirror or glide planes.
Figure~\ref{fig:degeneracy_valence} shows the color maps of the energy difference between the two highest valence bands and between the two lowest conduction bands of \YCI.
The calculations are performed in the mirror invariant plane $k_x = 0$ and the glide invariant plane $k_y = 0$.
All four panels show the degeneracy of the bands forming 2-dimensional irreps on the $k_z$ axis.
While no additional band degeneracy is found in the $k_y = 0$ plane  [Figs.~\ref{fig:degeneracy_valence}c and d], the two highest valence bands [Fig.~\ref{fig:degeneracy_valence}a] and two lowest conduction bands [Fig.~\ref{fig:degeneracy_valence}b] of \YCI have nodal lines in the $k_x=0$ plane extending from the Dirac point ($k_z^\mathrm{D} \simeq \SI{0.237}{\angstrom^{-1}}$) to the diagonal directions, $k_z>k_z^\mathrm{D}$ in a and $k_z<k_z^\mathrm{D}$ in b.

These properties are explained using the effective $k \cdot p$ Hamiltonian expanded around the Dirac point up to the second order in $k$. 
The effective Hamiltonian has the form
\begin{align}
\label{eq:kp_C6v}
H(\boldrm{k}) &= c_1 k_z \Gamma_{0,0} + c_2 k_z \Gamma_{3,0} + c_3 (k_x \Gamma_{2,0} - k_y \Gamma_{1,2}) \nonumber\\
&\phantom{{}={}} + c_4 (k_x \Gamma_{1,0} + k_y \Gamma_{2,2}) + c_5 (k_x^2 + k_y^2) \Gamma_{0,0} \nonumber\\
&\phantom{{}={}} + c_6 k_z^2 \Gamma_{0,0} + c_7 [(k_x^2 - k_y^2) \Gamma_{3,3} - 2 k_x k_y \Gamma_{0,1}] \nonumber\\
&\phantom{{}={}} + c_8 (k_x k_z \Gamma_{2,0} - k_y k_z \Gamma_{1,2}) + c_9 (k_x k_z \Gamma_{1,0} + k_y k_z \Gamma_{2,2}) \nonumber\\
&\phantom{{}={}} + c_{10} (k_x^2 + k_y^2) \Gamma_{3,0} + c_{11} k_z^2 \Gamma_{3,0} \nonumber\\
&\phantom{{}={}} + c_{12}  [(k_x^2 - k_y^2) \Gamma_{0,3} - 2 k_x k_y \Gamma_{3,1}], 
\end{align}
where the $\Gamma$ matrices are the direct products of the Pauli matrices $\sigma_i$ ($i=1,2,3$) and a $2\times2$ unit matrix $\sigma_0$, i.e.  $\Gamma_{i,j} = \sigma_i \otimes \sigma_j$.
This Hamiltonian has a sixfold rotational symmetry around the $z$ axis
$D(C_{6z}) H(\boldrm{k}) D(C_{6z})^\dagger = H(C_{6z} \boldrm{k})$
[$D(C_{6z}) = -(\Gamma_{3,0} + \mathrm{i} \sqrt{3}\Gamma_{0,2})/2$],
and a mirror symmetry about the $zx$ plane
$D(M_y) H(\boldrm{k}) D(M_y)^\dagger = H(M_y \boldrm{k})$
[$D(M_y) = \Gamma_{0,3}$],
hence it has $C_{6v}$ symmetry.
Note that a previous study~\cite{YU2022375} has also systematically constructed the $k \cdot p$ theory in \emph{linear order} in $k$, whereas our Hamiltonian in Eq.~\eqref{eq:kp_C6v} takes into account the \textit{second-order} terms that are crucial for the following discussions.
The band structure obtained by first-principles calculations is well fitted in the neighborhood of the Dirac point by the Hamiltonian in Eq.~(\ref{eq:kp_C6v}), and the obtained parameters are presented in Supplementary Note 4.
The eigenvalues of $H(\boldrm{k})$ in the $k_x = 0$ plane are
\begin{align}
E(0, k_y, k_z) &= [c_5 + (-1)^{s_1} c_7] k_y^2 + (c_1 + c_6 k_z) k_z \nonumber\\
&\phantom{{}={}} + (-1)^{s_2} \biggl\{ (c_3 + c_8 k_z)^2 k_y^2 + (c_4 + c_9 k_z)^2 k_y^2 \nonumber\\
&\phantom{{}={}} \left. + \left[c_2 k_z + c_{10} k_y^2 + (-1)^{s_1} c_{12} k_y^2 + c_{11} k_z^2\right]^2 \right\}^{1/2},
\end{align}
where $s_1, s_2 \in \{0, 1\}$.
The valence bands ($s_2=1$) are degenerate at $k_y \ne 0$, when the condition
\begin{equation}
\label{eq:nodal-line_valence}
c_7 \sqrt{(c_3^2 + c_4^2) k_y^2 + c_2^2 k_z^2} = c_2 c_{12} k_z
\end{equation}
is satisfied in the leading order of $k$.
By squaring both sides we obtain
\begin{equation}
\label{eq:nodal-line_valence_squared}
c_7^2 (c_3^2 + c_4^2) k_y^2 = c_2^2 (c_{12}^2 - c_7^2) k_z^2.
\end{equation}
Therefore, when $\abs{c_{12}} > \abs{c_7}$, the valence bands are degenerate along the straight lines $k_z\propto\pm k_y$ emanating from the Dirac point.
Note that, according to Eq.~(\ref{eq:nodal-line_valence}), the degeneracies occur only in the part of the lines described by Eq.~(\ref{eq:nodal-line_valence_squared}) that satisfies $c_2 c_7 c_{12} k_z > 0$.
Similarly, when $\abs{c_{12}} > \abs{c_7}$, the conduction bands are degenerate in the part of the lines represented by Eq.~(\ref{eq:nodal-line_valence_squared}) that satisfies $c_2 c_7 c_{12} k_z < 0$.
On the other hand, when $\abs{c_7} > \abs{c_{12}}$, the valence bands and the conduction bands on the $k_y = 0$ plane are degenerate in the part of the lines represented by
\begin{equation}
\label{eq:nodal-line_valence_squared_2}
c_{12}^2 (c_3^2 + c_4^2) k_x^2 = c_2^2 (c_7^2 - c_{12}^2) k_z^2
\end{equation}
that satisfies $c_2 c_7 c_{12} k_z > 0$ and $c_2 c_7 c_{12} k_z < 0$, respectively.
We note that $0<c_2<c_7<c_{12}$ in \YCI.
From the above discussion, we conclude that in the spinless Dirac semimetal phase protected by $C_{6v}$ symmetry, both the valence bands and the conduction bands are degenerate along the straight lines passing through the Dirac point extending to the opposite directions to each other on one of the two equivalent mirror invariant planes determined by the relative magnitude of $c_7$ and $c_{12}$.

\subsection{Surface states}

\begin{figure}[thp]
\includegraphics[width=8cm]{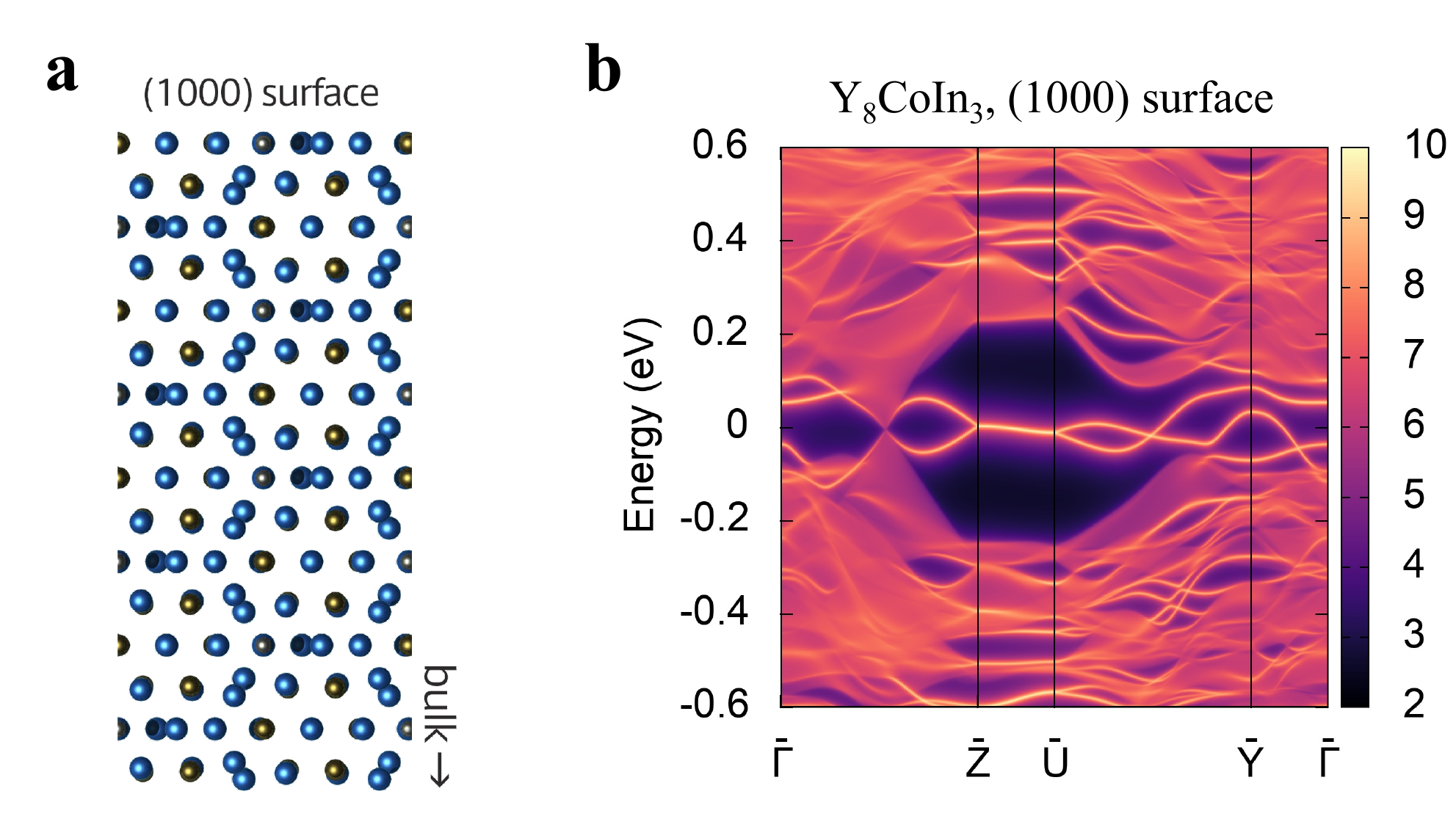}
\caption{\label{fig:Surface}
\textbf{Surface states in \ce{Y8CoIn3}.}
\textbf{a} The surface termination used in the calculation.
\textbf{b} The surface states for the (1000) surface.
}
\end{figure}

We calculate the surface states for the (1000) surface of \YCI, using the surface termination shown in Fig.~\ref{fig:Surface}a and the surface BZ shown in Fig~\ref{fig:SpinlessBand}b.
As shown in Fig.~\ref{fig:Surface}b, two nontrivial surface bands emerge from the projection of the bulk Dirac point.
The absence or presence of the nontrivial surface bands in the $\bar{\Gamma}$-$\bar{Z}$ line corresponds to the Zak phase values of 0 or $\pi$, respectively. 
Also, the number of nontrivial surface bands corresponds to the number of glide sectors for which the Zak phase is $\pi$.
This correspondence between the Zak phases for each sector and the nontrivial surface bands is reproduced by the $4 \times 4$ tight-binding model with the $C_{6v}$ point group symmetry and time reversal symmetry, as described in Supplementary Note 5.

The crystal maintains the glide symmetry with respect to the $zx$ plane when the surface is in this direction.
The $\bar{Z}$-$\bar{U}$ line is invariant up to the reciprocal lattice vector under the product of the glide operation $\{M_y|00\frac{1}{2}\}$ and the time reversal operation $\Theta$.
The product $\{M_y|00\frac{1}{2}\}\Theta$ is an anti-unitary operator and $(\{M_y|00\frac{1}{2}\}\Theta)^2$ is $-1$ when $ck_z = \pi$; therefore the Kramers-like degeneracy occurs on the $\bar{Z}$-$\bar{U}$ line.
This explains the degeneracy of the two midgap surface states along the $\bar{Z}$-$\bar{U}$ line.

\subsection{Topological transitions}

The spinless Dirac semimetallic phases are located at the phase boundaries of various topological phases.
By applying uniaxial pressure to \YCI, we can transform it into a multi-band nodal-line or Weyl semimetallic phases characterized by the second SW class.
Additionally, the use of lanthanides as substitutes can introduce magnetism and strong SOC since \YCI mainly consists of Y.

\subsubsection{The second Stiefel-Whitney class}

\begin{figure*}[htp]
\includegraphics[width=17cm]{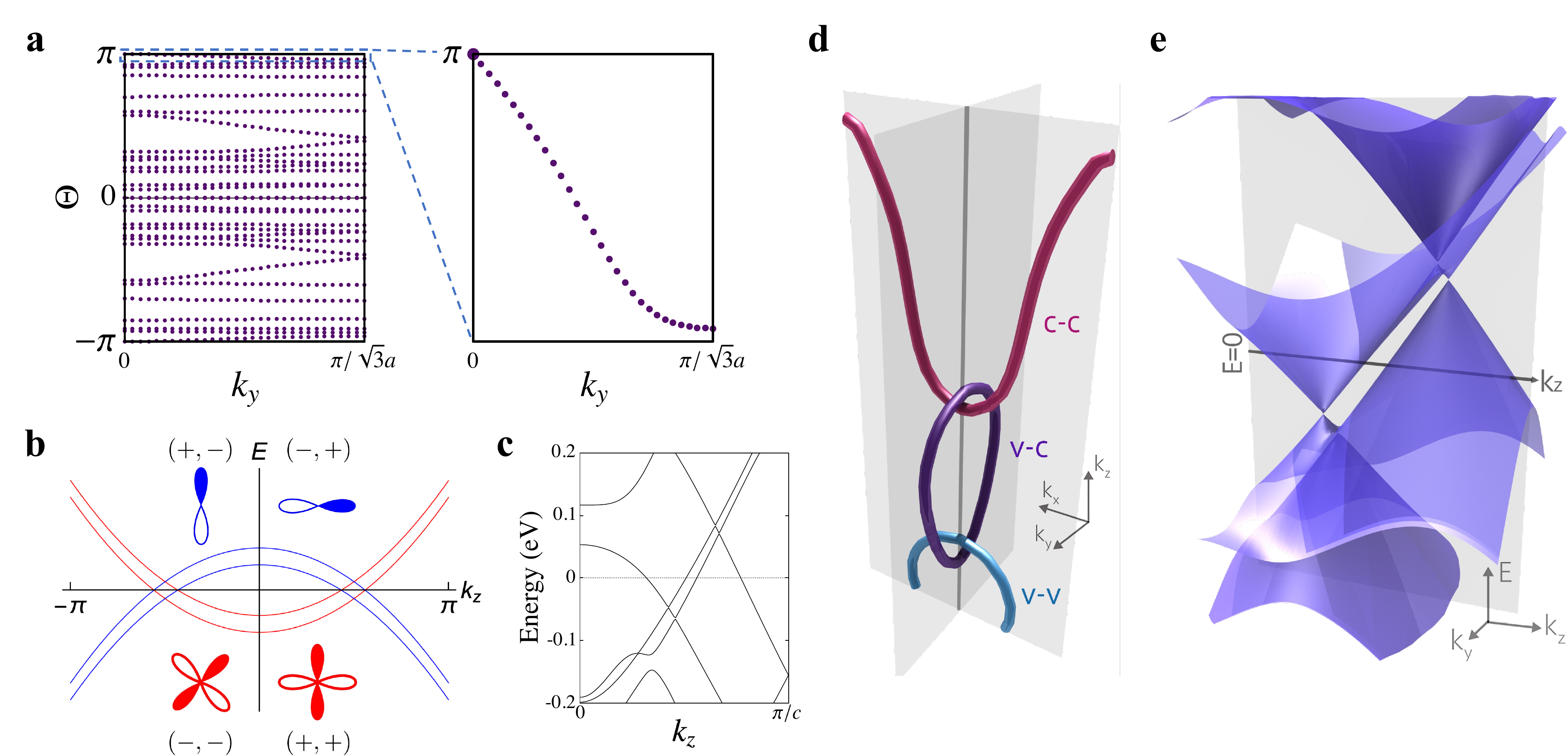}
\caption{\label{fig:w2}
\textbf{The second SW class and the topological phase transitions of Y$_8$CoIn$_3$.}
\textbf{a} The Wilson loop spectrum of \YCI calculated in the $k_z = 0$ plane. The integral path of the Wilson loop is along the $k_x$ direction at fixed $k_y$.
\textbf{b} Schematic band structure of $C_{2z}$ and $\Theta$ symmetric systems with different second SW classes in the $k_z = 0$ and $k_z = \pi$ planes.
The colors of the bands, red and blue, correspond to the $C_{2z}$ eigenvalue values, $+$ and $-$, respectively.
The atomic orbital-like bases and their mirror eigenvalues (eigenvalues for $M_x$ and for $M_y$ in this order) are also written for the case of $C_{2v}$ symmetry.
In this case, the degeneracies between a red band and a blue band occur on either the $k_x=0$ plane or the $k_y=0$ plane.
\textbf{c} The band structure on the $k_z$ axis of \YCI compressed by $5\%$ in the $x$ direction.
\textbf{d} The nodal structure of \YCI compressed by $5\%$ in the $x$ direction.
The nodal line between the valence and conduction bands is drawn in purple, and the nodal line between the two highest valence bands (lowest conduction bands) is drawn in blue (red).
The gray planes represent the $k_x = 0$ and $k_y = 0$ planes.
\textbf{e} The 3-dimensional band structure in the $k_x = 0$ plane of \YCI compressed by $5\%$ in the direction making an angle of $\pi/4$ from the $x$ axis. 
The gray plane shows the $k_z$-$E$ plane at $k_y = 0$.
}
\end{figure*}

In the space group $P6_3mc$, the $k_z = 0$ plane is invariant under the product of the twofold screw operation $\{C_{2z}|00\frac{1}{2}\}$ and the time reversal operation $\Theta$.
Since $\{C_{2z}|00\frac{1}{2}\} \Theta$ is an anti-unitary operator and satisfies $(\{C_{2z}|00\frac{1}{2}\} \Theta)^2 = +1$ in the $k_z = 0$ plane, we can adopt the real gauge for the wavefunctions in the $k_z = 0$ plane.
Recalling that \YCI has a finite energy gap throughout the $k_z = 0$ plane, the Hamiltonian on this plane is topologically characterized by a $\mathbb{Z}_2$ invariant called 
the second SW class~\cite{PhysRevB.96.155105,PhysRevB.99.235125,PhysRevB.102.115135,PhysRevB.104.195114,PhysRevLett.121.106403,PhysRevX.9.021013}.
The second SW class $w_2$ can be obtained by the Wilson loop method~\cite{PhysRevB.83.235401,PhysRevB.84.075119,PhysRevB.100.195135,PhysRevLett.121.106403,PhysRevX.9.021013}.
Figure~\ref{fig:w2}a shows the Wilson loop spectrum of \YCI calculated in the $k_z = 0$ plane, where the integral path of the Wilson loop is along the $k_x$ direction at fixed $k_y$.
The spectrum has one linear crossing at $\pi$, indicating $w_2 = 1$ in this plane.
On the other hand, in the $k_z = \pi/c$ plane, $(\{C_{2z}|00\frac{1}{2}\} \Theta)^2$ is $-1$ instead of $+1$.
In this case, the second homotopy group of the corresponding classifying space is the trivial group and gapped Hamiltonians have no topological distinction~\cite{PhysRevB.96.155105}.

In fact, when the system has both $C_{2z}$ (or screw) and $\Theta$ symmetries, $w_2$ can be determined from the $C_{2z}$ eigenvalues as follows~\cite{PhysRevB.99.235125,PhysRevLett.121.106403}:
\begin{equation}
(-1)^{w_2} = \prod_{i=1}^{4} (-1)^{\lfloor N_{\mathrm{occ}}^{-}(\Gamma_i)/2 \rfloor},
\end{equation}
where $\{\Gamma_i\}$ are the $C_{2z}$ invariant points on the plane where $w_2$ is evaluated, $N_{\mathrm{occ}}^{-}(\Gamma_i)$ is the number of occupied bands with negative $C_{2z}$ eigenvalues at $\Gamma_i$, and $\lfloor\cdot\rfloor$ represents the floor function. 
From this formula, it can be seen that for symmorphic systems with $C_{2z}$ and $\Theta$ symmetries, when the $w_2$ indices in the $k_z = 0$ and $k_z = \pi$ planes are different, two band inversions occur between bands with different $C_{2z}$ eigenvalues as shown in Fig.~\ref{fig:w2}b, which is called a double band inversion.
In this case, four bands form a gap-closing object in $0 < k_z < \pi$, which 
mediates the two planes with different $w_2$.
In \YCI, the two 2-dimensional irreps $\Delta_5$ and $\Delta_6$, which are distinguished by the twofold screw eigenvalues, intersect on the $k_z$ axis, similarly to the band structure in Fig.~\ref{fig:w2}b.
Although $w_2$ cannot be defined on the $k_z=\pi$ plane in \YCI, the Dirac point can be considered to 
act as an intermediary between the $k_z=0$ plane hosting the nontrivial $w_2$ 
and the $k_z=\pi$ plane having the trivial topology.

The $C_{6v}$ symmetry is actually not required for the definition of $w_2$ and its calculation from the $C_{2z}$ eigenvalues.
We thus investigate the topological phase transitions of \YCI under uniaxial strains that break $C_{6v}$ symmetry but maintain $C_2$ symmetry.
Compressing the \YCI lattice by $5\%$ in the $x$ direction results in space group $Cmc2_1$ (No.~36), and its little co-group on the $k_z$ axis is $C_{2v}$.
Figure~\ref{fig:w2}c shows the band structure calculated on the $k_z$ axis, where the 2-dimensional irreps are split due to the symmetry reduction.
Since the twofold screw eigenvalues of the occupied bands in the $k_z=0$ plane remain unchanged, $w_2$ is invariant.
However, in this system, a nodal line on the mirror invariant plane $k_x=0$ 
mediates the $k_z = 0$ and $k_z = \pi$ planes instead of the Dirac point.
Furthermore, in this system, nodal lines between the two highest valence bands and between the two lowest conduction bands lie on the glide invariant plane $k_y = 0$ and pass inside the nodal line between the valence and conduction bands as shown in Fig.~\ref{fig:w2}d.
The fact that the nodal lines appear on the two different mirror-invariant planes can be understood from the symmetry of the bands.
For example, a band with the $d_{xy}$ orbital symmetry can intersect a band with the $p_x$ orbital symmetry in the $k_y = 0$ plane, whereas it can intersect a band with the $p_y$ orbital symmetry in the $k_x =0$ plane [see Fig.~\ref{fig:w2}b].
Notably, wavefunctions with positive (negative) $C_2$ eigenvalues have the same (different) eigenvalues for the two orthogonal mirror reflections.

We analyze this system using the $k \cdot p$ Hamiltonian in Eq.~(\ref{eq:kp_C6v}).
The $k$-independent perturbation that breaks $C_{6v}$ but preserves $C_{2v}$ symmetry has the form $c_{13}\Gamma_{0,3} + c_{14}\Gamma_{3,3}$.
In the following, we investigate the nodal lines near $\boldrm{k} = \boldrm{0}$ in the leading order of $k$, assuming that the perturbation is sufficiently small.
First, let $c_{13}=0$ for simplicity.
Then, the eigenvalues on the $k_x = 0$ plane are
\begin{align}
E(0, k_y, k_z) &={} -(-1)^{s_1} c_{14} + [c_5 + (-1)^{s_1} c_7] k_y^2 + (c_1 + c_6 k_z) k_z \nonumber\\
&\phantom{{}={}} + (-1)^{s_2} \biggl\{ (c_3 + c_8 k_z)^2 k_y^2 + (c_4 + c_9 k_z)^2 k_y^2 \nonumber\\
&\phantom{{}={}} \left. + \left[c_2 k_z + c_{10} k_y^2 + (-1)^{s_1} c_{12} k_y^2 + c_{11} k_z^2\right]^2 \right\}^{1/2},
\end{align}
where $s_1, s_2 \in \{0, 1\}$, and the top valence and bottom conduction bands are degenerate along the curve defined by
\begin{equation}
\label{eq:C2v_v-c_nodal-line_c14}
(c_3^2 + c_4^2 + 2 c_7 c_{14}) k_y^2 + c_2^2 k_z^2 = c_{14}^2.
\end{equation}
Since the perturbation $c_{14}$ is assumed to be small, this represents an ellipse around $\boldrm{k} = \boldrm{0}$.
We note that this nodal ring cannot be gapped out by tuning $c_{14}$.
Suppose we start from the nodal-line semimetal phase and change $c_{14}$ through 0.
As $c_{14}$ approaches 0, the size of the nodal ring decreases.
At $c_{14} = 0$ the nodal ring shrinks to the Dirac point, but when $c_{14}$ becomes finite after the sign reversal, it grows into a nodal ring again.

Similarly, on the $k_y = 0$ plane, the two valence bands and also the two conduction bands are degenerate along
\begin{equation}
[c_{10} \pm \sgn (c_{14}) c_{12}] k_x^2 + c_2 k_z = 0,
\label{eq: two parabolas}
\end{equation}
where $\pm$ corresponds to the valence and conduction bands, respectively, and $\sgn (\cdot)$ denotes the sign function.
The degenerate $\boldrm{k}$ points form two parabolas with the vertex at $\boldrm{k} = \boldrm{0}$ and thread the nodal line defined by Eq.~(\ref{eq:C2v_v-c_nodal-line_c14}).
These features coincide with those of nodal lines with $\mathbb{Z}_2$ monopole charge in $PT$-symmetric systems~\cite{PhysRevLett.121.106403,PhysRevB.92.081201}, where $PT$ is the product of space inversion and time reversal operators.
However, it should be noted that our system does not have space inversion symmetry, but has $C_{2v}$ symmetry instead.
In the general case with $c_{13} \neq 0$, the planes where the nodal lines emerge switch between the $k_x = 0$ and $k_y = 0$ planes depending on the relative magnitude of $c_{13}$ and $c_{14}$, and the two connected parabolas of Eq.~(\ref{eq: two parabolas}) are turned into two branches of a hyperbola as shown in Fig.~\ref{fig:w2}d.
The details of the calculations are presented in Supplementary Note 6.
Note that, depending on the sign of the perturbation parameters, the degeneracies between the valence bands and between the conduction bands represented by Eq.~(\ref{eq:nodal-line_valence_squared}) and Eq.~(\ref{eq:nodal-line_valence_squared_2}) can remain in the plane, but shift enough from the $k_z$ axis to move outside of the nodal ring in Eq.~(\ref{eq:C2v_v-c_nodal-line_c14}).

When the system is compressed in the direction with an angle of $\pi/4$ from the $x$ axis, the space group transforms into $P2_1$ (No.~4), and its little co-group on the $k_z$ axis becomes $C_2$.
Under this distortion, the $k_z = 0$ plane still preserves the nontrivial $w_2$.
However, the gap-closing object 
bridging the $k_z = 0$ and $k_z = \pi$ planes becomes a pair of Weyl points with opposite chirality [Fig.~\ref{fig:w2}e].
This phenomenon is typical of Dirac semimetal phases with broken spatial inversion symmetry and has been discussed in previous studies~\cite{PhysRevB.103.205151,PhysRevLett.121.106404,PhysRevResearch.4.033170,Zhang2023}.
The analytical argument with the $k \cdot p$ Hamiltonian is given in the Supplementary Note 7.

The results of the second SW class for the Dirac, nodal-line, and Weyl semimetals are combined as follows.
The gapped Hamiltonian in the $k_z = 0$ plane of nonsymmorphic systems with $\{C_{2z}|00\frac{1}{2}\}$ and $\Theta$ is characterized by the second SW class $w_2$, which takes the $\mathbb{Z}_2$ values.
If $w_2 = 1$ and the system has a finite energy gap except on the $k_z$ axis, a gap-closing object is formed by four bands on the $k_z$ axis.
Depending on the symmetry, it can be a pair of Weyl points, a nodal line, or a Dirac point.
This is also the case for symmorphic systems with different $w_2$ on the $k_z = 0$ and $k_z = \pi$ planes.
In this perspective, the Dirac semimetal such as \YCI can be understood as a system with a double band inversion where the parameters are fine-tuned by the symmetry.
\tcr{The surface states of the three topological semimetal phases characterized by the second SW class are discussed in Supplementary Note 8.}

\subsubsection{Spinful Dirac semimetal}
\label{subsubsec:Spinful}

\begin{figure}[htp]
\includegraphics[width=8cm]{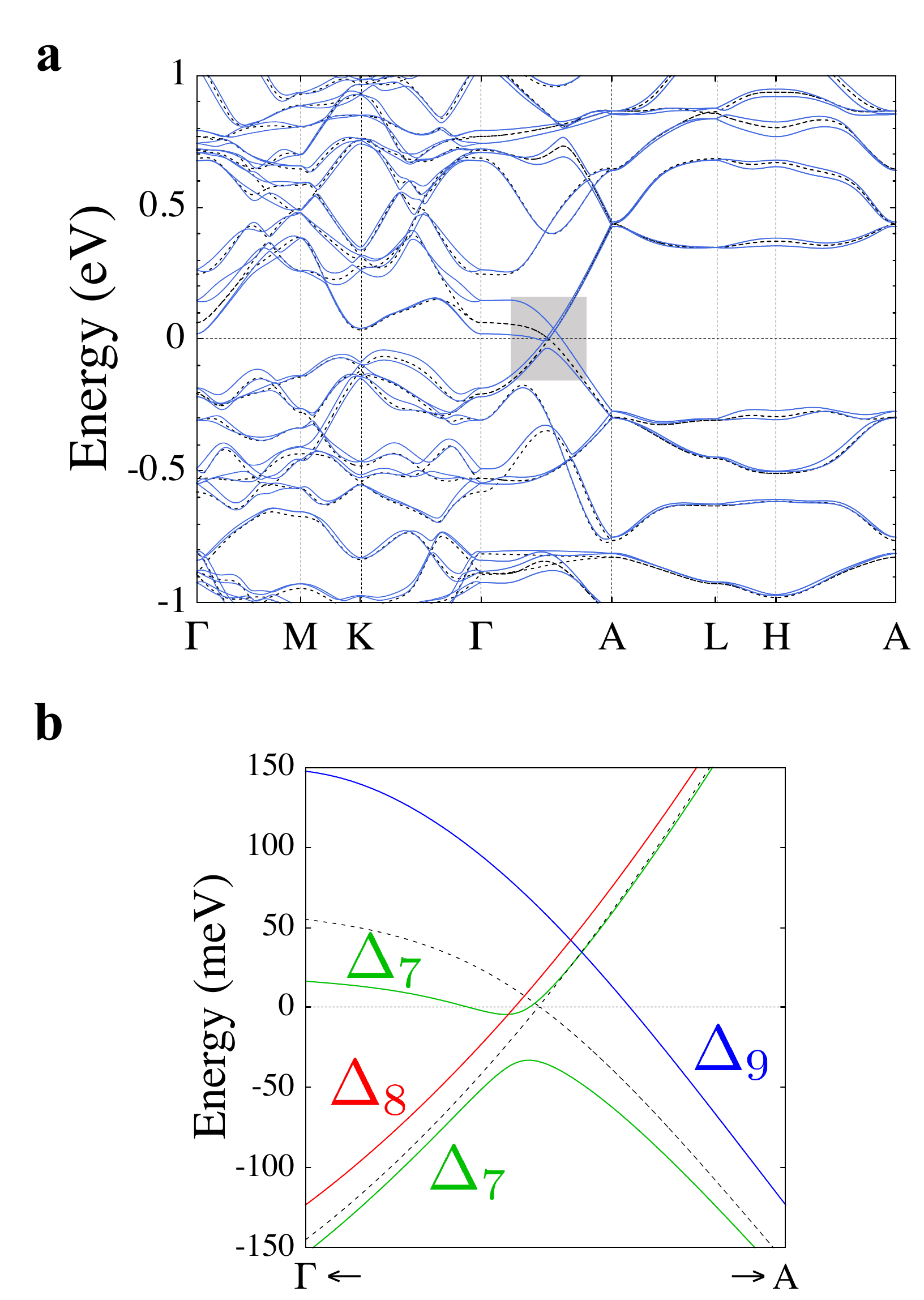}
\caption{\label{fig:SpinfulBand}
\textbf{The electronic band structure of Lu$_8$CoGa$_3$ calculated with and without the SOC.}
\textbf{a} The electronic band structure of Lu$_8$CoGa$_3$ calculated by the GGA.
The blue solid line and the black dashed line show the result with and without the SOC, respectively.
The energy is measured from the Fermi level in the spinless calculation.
\textbf{b} Magnified band structure in the shaded region in \textbf{a}.
The irreps of the bands with the SOC are indicated.
}
\end{figure}

Figure~\ref{fig:SpinfulBand}a shows the electronic band structure of Lu$_8$CoGa$_3$ calculated in the presence and absence of the SOC.
The Lu $5d$ orbitals have a strong SOC compared to the Y $4d$ orbitals, resulting in the significant SOC splitting near the Fermi level in Lu$_8$CoGa$_3$.
In Fig.~\ref{fig:SpinfulBand}b, the single-valued representations $\Delta_5$ and $\Delta_6$ transform into \tcr{$\Delta_7 \oplus \Delta_8$ and $\Delta_7 \oplus \Delta_9$}, respectively, when the SOC is turned on.
The double-valued representations $\Delta_7$, $\Delta_8$ and $\Delta_9$ are all 2-dimensional irreps that are distinguished by the screw symmetry around the $z$ axis.
The two bands belonging to $\Delta_7$ hybridize and open the energy gap. 
The hybridization of bands belonging to different representations is prohibited;
therefore, three Dirac points are formed near the Fermi level, which are not related by any symmetry and have different energies.
The SOC gap of Lu$_8$CoGa$_3$ can be measured by taking the energy difference between the maximum of the lower $\Delta_7$ band and the intersection of the $\Delta_8$ and $\Delta_9$ bands, which approximately equals \SI{75}{\meV}.
When Lu is replaced with Y, the magnitude of the SOC is drastically reduced to $1/3$ or less (see Supplementary Note 3).

\subsubsection{Magnetic Weyl semimetal}
\begin{figure*}[htp]
\includegraphics[width=16cm]{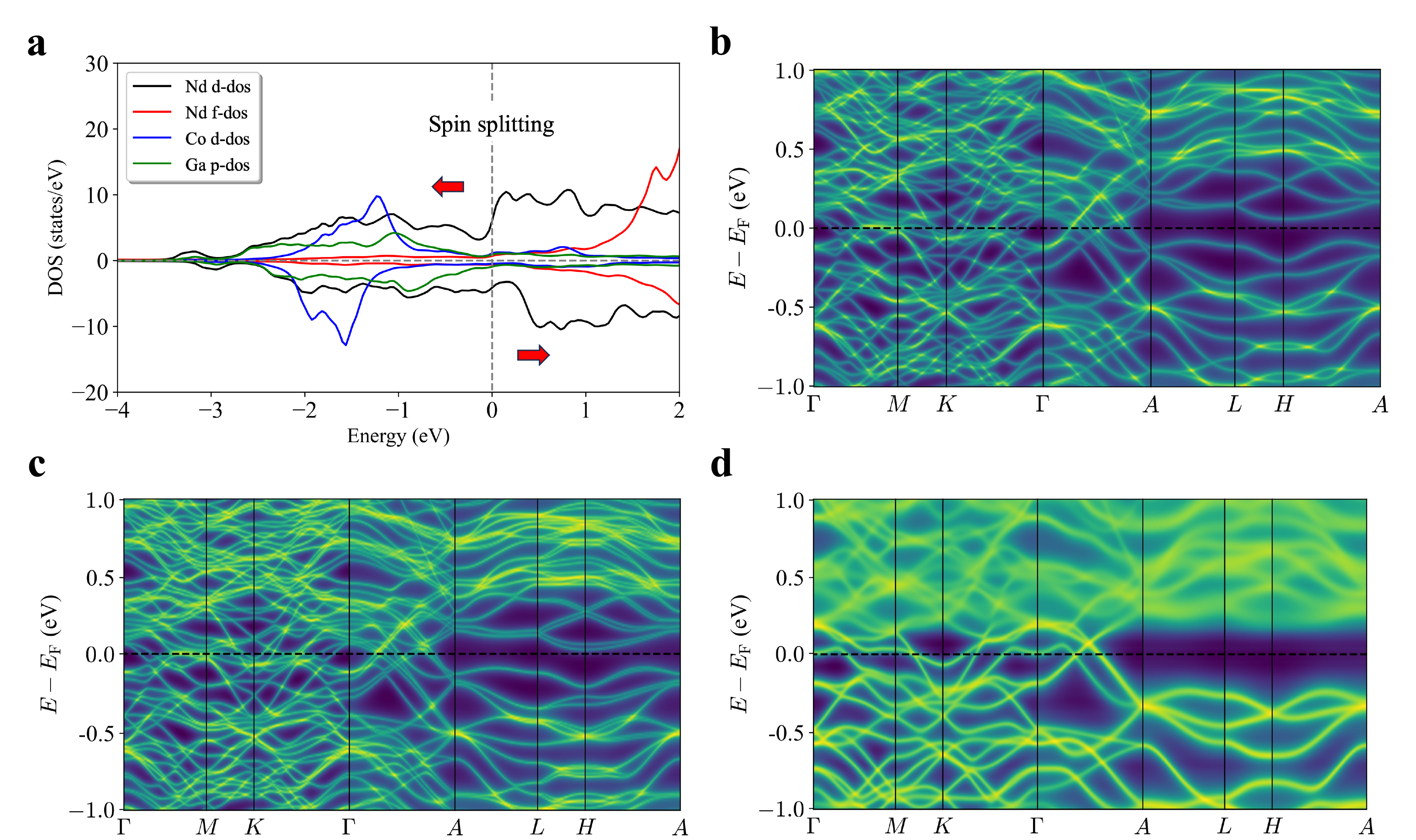}
\caption{
\textbf{The electronic structure of the magnetic material Nd$_8$CoGa$_3$ calculated by the KKR method.}
\textbf{a} The PDOS of Nd$_8$CoGa$_3$ calculated without the SOC.
\textbf{b} The electronic band structure of Nd$_8$CoGa$_3$ without the SOC.
\textbf{c} The electronic band structure of Nd$_8$CoGa$_3$ with the SOC included self-consistently in ferromagnetic configurations.
\textbf{d} The electronic band structure of Nd$_8$CoGa$_3$ in the paramagnetic state computed using the DLM approach.
The smearing of the bands is due to the finite temperature effects.
}
\label{fig:magnetic}
\end{figure*}

In the presence of magnetic rare-earth atoms, the time-reversal symmetry is broken. 
It is intriguing to examine the evolution of topological features near the Fermi energy in the presence of a finite magnetization.
The Dirac cone is expected to split into a pair of Weyl points with opposing chiralities~\cite{RevModPhys.90.015001,PhysRevB.85.195320,PhysRevB.88.125427}.
We have opted for Nd$_{8}$CoGa$_{3}$ as our magnetic prototype system, applying an all-electron full-potential KKR Green function method with the GGA for exchange-correlation energy~\cite{Perdew91}. 
The nonmagnetic La$_8$CoGa$_3$ has a small Fermi surface around the $k_z=0$ plane (see Supplementary Note 3), and there is also a low DOS at the Fermi level when Nd replaces La [Fig.~\ref{fig:magnetic}a].
The Nd magnetic atom is in a $3+$ state, and has a magnetic moment of $3.19\,\mu_{\text{B}}$ per atom in agreement with Hund's rules (the moment points along the $c$-axis). 
Due to the finite spin polarization on Nd, an induced moment of $0.43\,\mu_\text{B}$ emerges on the Co atoms with an antiferromagnetic coupling to the Nd atoms. 
Additionally, a weak induced moment of $0.04\,\mu_\text{B}$ appears on the Ga atoms.

The electronic band structure for Nd$_{8}$CoGa$_{3}$ without the SOC is shown in Fig.~\ref{fig:magnetic}b. 
As expected from the density of states, the gap is populated by the Co $3d$ and Ga $4p$ bands, as a consequence of the spin splitting introduced by the Nd magnetic moment [see Fig.~\ref{fig:magnetic}a]. 
Nonetheless, the gap remains open and unaltered along the [AL, LH, HA] directions.
This is because a large energy gap is opened around the $k_z=\pi$ plane in \RCX materials, and the number of occupied bands in the magnetic state, excluding the $4f$ orbitals, is unchanged compared to the nonmagnetic state.
\tcr{The band structure for each spin component is presented in Supplementary Note 9.}
Each of the Dirac points in the $\Gamma$-A line splits into a pair of Weyl points in the presence of the SOC as shown in Fig.~\ref{fig:magnetic}c.
Furthermore, given that the filling of the gap originates from the induced spin polarization on Co and Ga, we investigate the electronic band structure in the paramagnetic phase. 
To simulate the paramagnetic state, we employ the disordered local moment (DLM) approach~\cite{gyorffy1985}, the fully magnetic disordered state is then established using the coherent potential approximation (CPA)~\cite{Gyorffy1972}. 
In the disordered state, the induced moments on the Co and Ga vanish, hence fewer bands are observed in the gap in comparison with the ferromagnetic state [see Fig.~\ref{fig:magnetic}d]. 
The magnetic disorder causes the smearing of the electronic bands. 
The dependence of bands populating the gap on the induced spin polarization indicates that the Fermi surface, and semimetallic character can be tuned via the magnetic order and finite temperature effects. 
\tcr{Since the spin splitting of itinerant $4d$ electrons is small, the band structure shown in Fig.~\ref{fig:magnetic}d is in good agreement with the band structure calculated with a pseudopotential that treats the Nd $4f$ orbitals as core states (see Supplementary Note 9).}
Lastly, as the magnetic order alters the topological band structure of Nd$_{8}$CoGa$_{3}$, we performed a preliminary analysis of the magnetic interactions in real space using the infinitesimal rotation method~\cite{liechtenstein1984}, and find that the intralayer and interlayer magnetic coupling is predominantly ferromagnetic with a transition temperature of $T_{\text{c}}=\SI{40}{\kelvin}$.
The relatively high temperature can be attributed to the high number of rare-earth atoms within the unit cell.

\section{CONCLUSION}
In summary, the ternary compounds $RE_8\mathrm{Co}X_3$ ($RE$ = rare earth elements, $X$ = Al, Ga, or In) are the ideal platform of the spinless Dirac semimetal and the topological transitions.
The strength of the SOC in these materials can be adjusted by substituting elements, and those containing Y have a smaller SOC.
The Dirac points in these materials are manifested by accidental degeneracy of the two different 2-dimensional irreps of $C_{6v}$ on the $k_z$ axis.
The degeneracies between the valence bands and between the conduction bands are generally lifted away from the $k_z$ axis, but there are degenerate lines on the mirror planes extending from the Dirac points.
The nontrivial topology of the bulk states is characterized by the Zak phases for each glide sector, which corresponds to the two midgap bands in the surface extending from the projection of the bulk Dirac point.
Furthermore, we discuss the topological phase transitions under the distortions and the elemental substitutions.
Uniaxial compression, which preserves the twofold screw symmetry, breaks the $C_{6v}$ symmetry and turns the Dirac point into the nodal line or the pair of Weyl points.
The magnetic lanthanide elements can replace Y, resulting in a spinful magnetic Weyl semimetallic phase in the system.

\section{METHOD}

The electronic structure of \YCI is calculated based on density functional theory (DFT) as implemented in the Vienna Ab initio Simulation Package (VASP)~\cite{PhysRevB.54.11169}.
For the exchange-correlation functional, we employ the GGA of Perdew, Burke, and Ernzerhof (PBE)~\cite{PhysRevLett.77.3865}.
The cutoff energy for the plane wave expansion is \SI{350}{\eV} and a $k$-point mesh of $8 \times 8 \times 12$ is adopted.
The calculations are based on the experimental crystal structure.
\tcr{The irreps of the bands are obtained using the IrRep package~\cite{IRAOLA2022108226,Elcoro:ks5574}.}
To calculate the PDOS, Zak phases, and surface states, we construct a tight-binding Hamiltonian from the DFT results using the Wannier90 package~\cite{Pizzi_2020}. 
From the Bloch functions within the range from \SI{4}{\eV} below to \SI{10}{\eV} above the Fermi level, we construct 114 Wannier functions for the Y $4d$ orbitals, the Co $3d$ orbitals, the In $5p$ orbitals, and the interstitial $s$ orbitals near the Co atoms.
The surface states are calculated by the iterative Green's function method~\cite{PhysRevB.28.4397, MPLopezSancho_1984, MPLopezSancho_1985} implemented in the WannierTools package~\cite{WU2017}.
The WannierTools package is also used, \textit{mutatis mutandis}, to calculate the Zak phases.
When we investigate topological phase transitions caused by uniaxial strains, the lattice is compressed in one direction and expanded uniformly in the two orthogonal directions so that the volume of the unit cell is preserved, and then the structure is optimized using the revised PBE for solids (PBEsol)~\cite{PhysRevLett.100.136406} as the exchange-correlation functional.

For the magnetic Weyl semimetals our first-principles simulations employ the all-electron full-potential KKR Green function method~\cite{papanikolaou2002} with and without SOC. 
The $4f$ electrons are treated using the DFT+$U$ approach in the Dudarev formulation~\cite{Dudarev2019} with $U=\SI{6}{\eV}$. 
The self-consistent energy contour is divided into 48 energy points with a $25 \times 25 \times 25$ $k$-mesh in the BZ. 
The electronic band structure along the high symmetry path represents the quasiparticle density of states (QDOS) computed directly from Green function in the ordered ferromagnetic or paramagnetic (DLM) states~\cite{ebert2011}.

\begin{acknowledgments}
The authors thank S. Murakami for the valuable discussions.
This work was supported by JST CREST (Grant No. JPMJCR19T2).
M.S. was supported by the Program for Leading Graduate Schools (MERIT-WINGS).
S.S. acknowledges financial support from JSPS KAKENHI Grants No. 23K13056 and No. 23K03333.
S.K. acknowledges financial support from JSPS KAKENHI Grants No. 19K14612 and No. 22K03478.
M.H. acknowledges financial support from PRESTO, JST (JPMJPR21Q6) and JSPS KAKENHI Grants No. 18H03678.
\end{acknowledgments}

\appendix{}

\section*{Supplementary Note 1. Classification of topological semimetals with fourfold degeneracy}

Here we discuss the classification of topological semimetal phases with quadruple degeneracy.
Supplementary Table~\ref{tab:classification_toposemi_fourfold} shows the classification according to the presence or absence of the SOC and the dimension of the nodes between valence and conduction bands.
Note that, as mentioned in the main text, we ignore the spin degrees of freedom when counting the degree of degeneracy in spinless systems.
The representation theory of space groups tells us which space groups can host each type of quadruple degeneracy.
As mentioned in the main text, there are two classes of a fourfold degenerate point, or a Dirac point, for both spinless and spinful systems.
In the presence of appropriate nonsymmorphic symmetry, a fourfold degenerate nodal line can exist in both spinless and spinful systems.

\begin{supptable*}[htp]
\caption{
\textbf{Classification of symmetry-protected topological semimetals with fourfold degeneracy.}
We ignore the spin degrees of freedom when counting the degree of degeneracy in spinless systems.
}
\label{tab:classification_toposemi_fourfold}
\renewcommand{\arraystretch}{1.25}
\begin{tabular}{|c||c|c|}
\hline
 & \multicolumn{2}{c|}{Dimension of nodes} \\
\cline{2-3}
 & 0D (point) & 1D (line) \\
\hline \hline 
\begin{tabular}{c} spinless \\ (with only orbital degrees of freedom) \end{tabular} & \hspace{1pt} \YCI (\textbf{This work}) \hspace{1pt} & \begin{tabular}{c} \ce{Al3FeSi2}~\cite{PhysRevB.96.155206} \\ \ce{LiBH}~\cite{Zhou_2020} \end{tabular} \\
\hline
\begin{tabular}{c} spinful \\ (with orbital and spin degrees of freedom) \end{tabular} & \begin{tabular}{c} \ce{Na3Bi}~\cite{Na3Bi_exp,PhysRevB.85.195320} \\ \ce{Cd3As2}~\cite{Neupane2014,Liu2014,Jeon2014,PhysRevLett.113.027603,PhysRevB.88.125427} \end{tabular} & \hspace{1pt} \ce{SrIrO3}~\cite{PhysRevB.85.115105,PhysRevB.86.085149,Chen2015} \hspace{1pt} \\
\hline
\end{tabular} 
\renewcommand{\arraystretch}{1.0}
\end{supptable*}

\section*{Supplementary Note 2. Band structure of $\bm{\mathrm{Y}_8\mathrm{CoIn}_3}$ obtained by the KKR method}

Supplementary Figure~\ref{fig:Y8CoIn3_KKR} shows the band structure of \YCI calculated by the KKR method in the absence of the SOC.
This is in good agreement with Fig.~2a in the main text.

\begin{suppfigure}[htp]
\includegraphics[width=8cm]{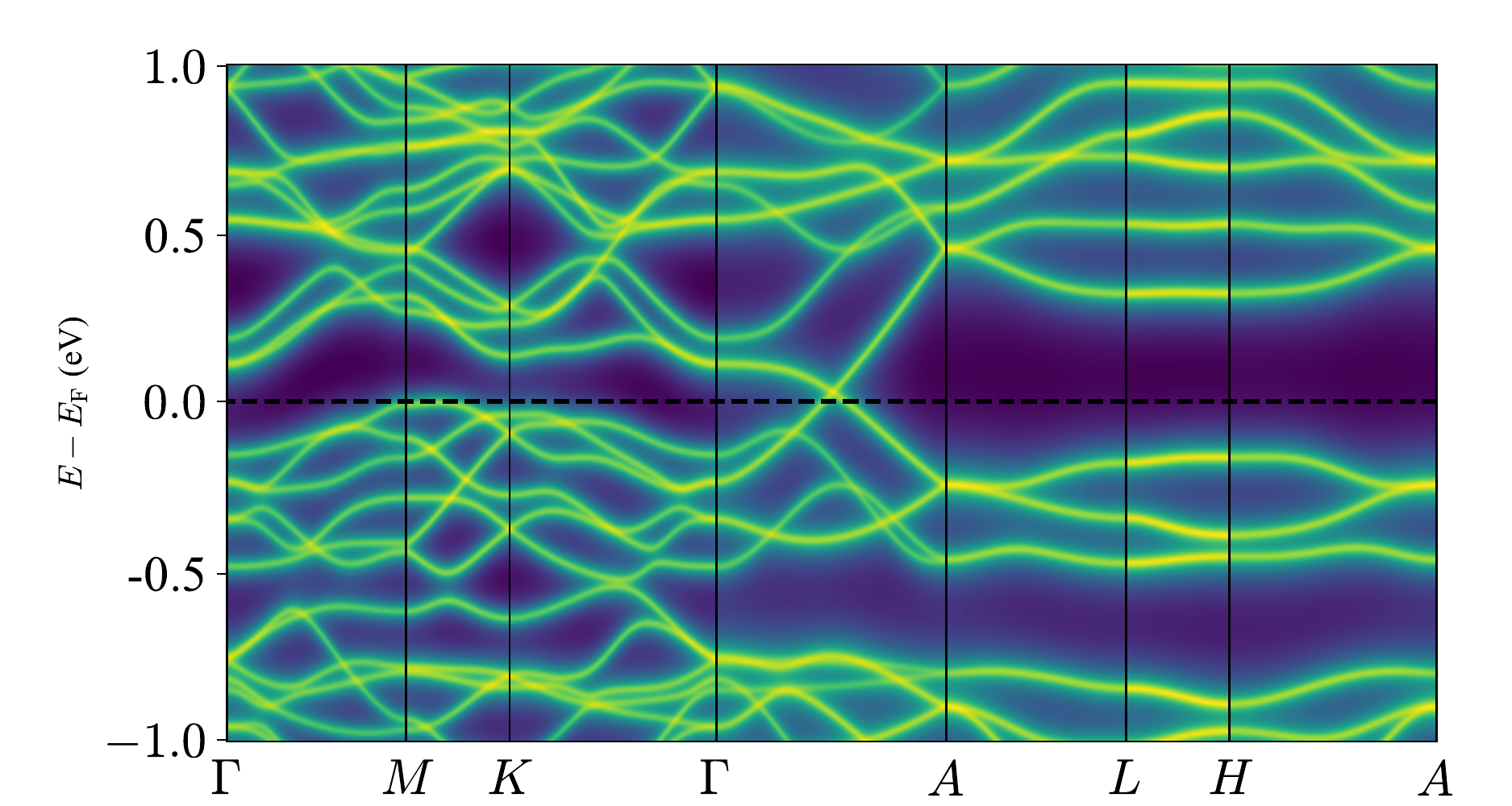}
\caption{\label{fig:Y8CoIn3_KKR}
\textbf{Band structure of $\bm{\mathrm{Y}_8\mathrm{CoIn}_3}$ obtained by the KKR method.}
The calculation is performed without the SOC.
}
\end{suppfigure}

\section*{Supplementary Note 3. Band structures of other materials in \RCX}
\label{sec:OtherMaterials}

\begin{suppfigure*}[htp]
\includegraphics[width=17cm]{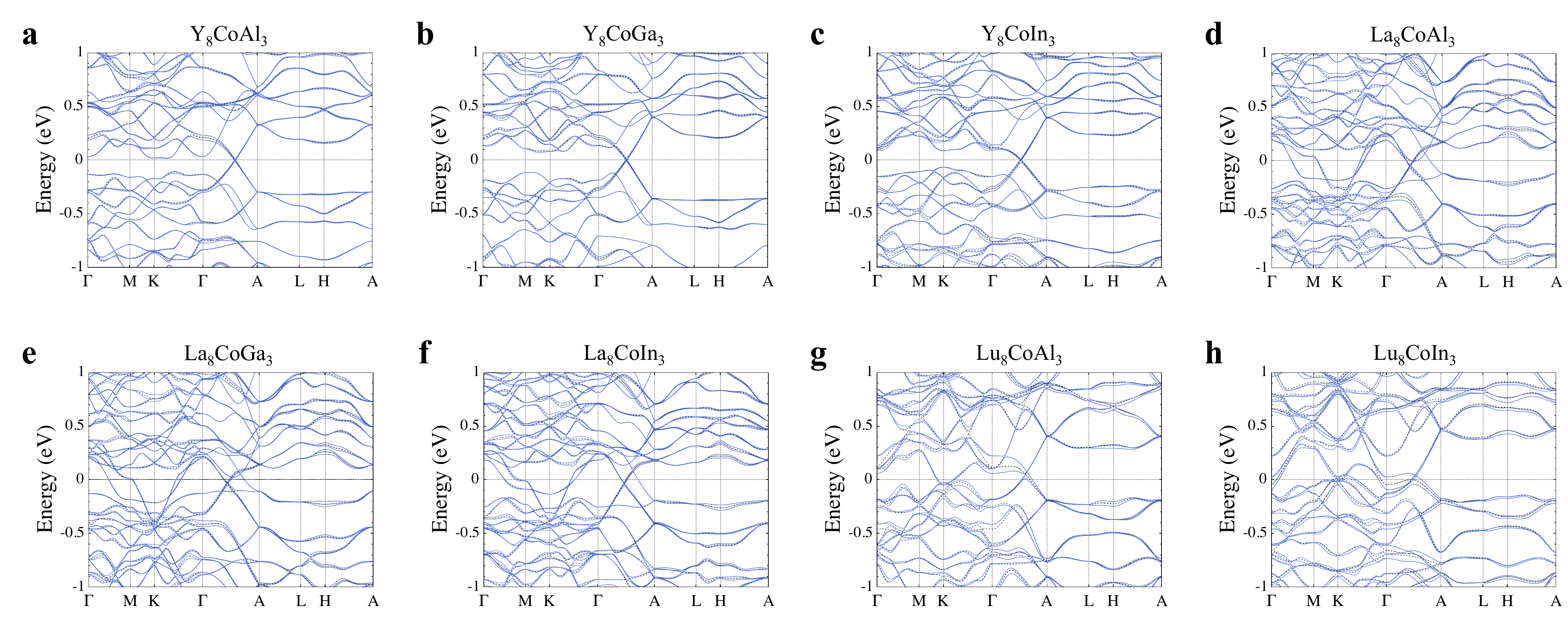}
\caption{\label{fig:OtherMaterialsBand}
\textbf{The electronic band structures of $\bm{RE_8\mathrm{Co}X_3}$ with $\bm{RE = \{\mathrm{Y}, \mathrm{La}, \mathrm{Lu}\}}$ and $\bm{X = \{\mathrm{Al}, \mathrm{Ga}, \mathrm{In}\}}$ other than $\bm{\mathrm{Lu}_8\mathrm{CoGa}_3}$, calculated with and without the SOC.}
The blue solid line and the black dashed line show the result with and without the SOC, respectively.
The energy is measured from the Fermi level.
\textbf{a} $\mathrm{Y}_8\mathrm{CoAl}_3$.
\textbf{b} $\mathrm{Y}_8\mathrm{CoGa}_3$.
\textbf{c} $\mathrm{Y}_8\mathrm{CoIn}_3$.
\textbf{d} $\mathrm{La}_8\mathrm{CoAl}_3$.
\textbf{e} $\mathrm{La}_8\mathrm{CoGa}_3$.
\textbf{f} $\mathrm{La}_8\mathrm{CoIn}_3$.
\textbf{g} $\mathrm{Lu}_8\mathrm{CoAl}_3$.
\textbf{h} $\mathrm{Lu}_8\mathrm{CoIn}_3$.
}
\end{suppfigure*}

In this section, we discuss the band structures of nonmagnetic materials in \RCX: \RCX with $RE = \{\mathrm{Y}, \mathrm{La}, \mathrm{Lu}\}$ and $X = \{\mathrm{Al}, \mathrm{Ga}, \mathrm{In}\}$ other than $\mathrm{Lu}_8\mathrm{CoGa}_3$.
In the calculations, the experimental crystal structure is used for $\mathrm{Lu}_8\mathrm{CoIn}_3$~\cite{DZEVENKO201314}, while for the other materials the crystal structures are optimized using 
the PBEsol in the GGA as the exchange-correlation functional.
The other calculation conditions are the same as for the \YCI calculation.
Supplementary Figure~\ref{fig:OtherMaterialsBand} shows the band structures calculated in the presence and absence of the SOC.
$\mathrm{Y}_8\mathrm{CoAl}_3$, $\mathrm{Y}_8\mathrm{CoGa}_3$ and $\mathrm{Y}_8\mathrm{CoIn}_3$ are ideal spinless Dirac semimetals, while the other materials have extra Fermi surfaces.
The SOC gaps of $\mathrm{Y}_8\mathrm{Co}X_3$ ($X$ = Al, Ga, In), measured in the same way as $\mathrm{Lu}_8\mathrm{CoGa}_3$ in the main text, are \SI{21}{\meV}, \SI{20}{\meV}, and \SI{28}{\meV}, respectively.

\section*{Supplementary Note 4. parameters of the $k\cdot p$ Hamiltonian}
\label{sec:kp_parameters}

\begin{supptable}[htp]
\caption{
\textbf{The fitted parameters for the Hamiltonian in Eq.(2) of the main text.}
}
\label{tab:kp_fitting_parameter}
\begin{tabular}{cccc}
\hline\hline
$c_1$ & $c_2$ & $c_3$ & $c_4$ \\
\hline 
$\SI{0.230}{\eV \angstrom}$ & $\SI{1.39}{\eV \angstrom}$ & $\SI{1.15}{\eV \angstrom}$ & $\SI{1.15}{\eV \angstrom}$ \\
$c_5$ & $c_6$ & $c_7$ & $c_8$ \\
\hline
$\SI{-0.732}{\eV \angstrom^2}$ & $\SI{-0.136}{\eV \angstrom^2}$ & $\SI{1.93}{\eV \angstrom^2}$ & $\SI{-1.91}{\eV \angstrom^2}$ \\
$c_9$ & $c_{10}$ & $c_{11}$ & $c_{12}$ \\
\hline
$\SI{0.615}{\eV \angstrom^2}$ & $\SI{0}{\eV \angstrom^2}$ & $\SI{1.93}{\eV \angstrom^2}$ & $\SI{2.65}{\eV \angstrom^2}$ \\
\hline\hline
\end{tabular} 
\end{supptable} 

\begin{suppfigure}[htp]
\includegraphics[width=7.5cm]{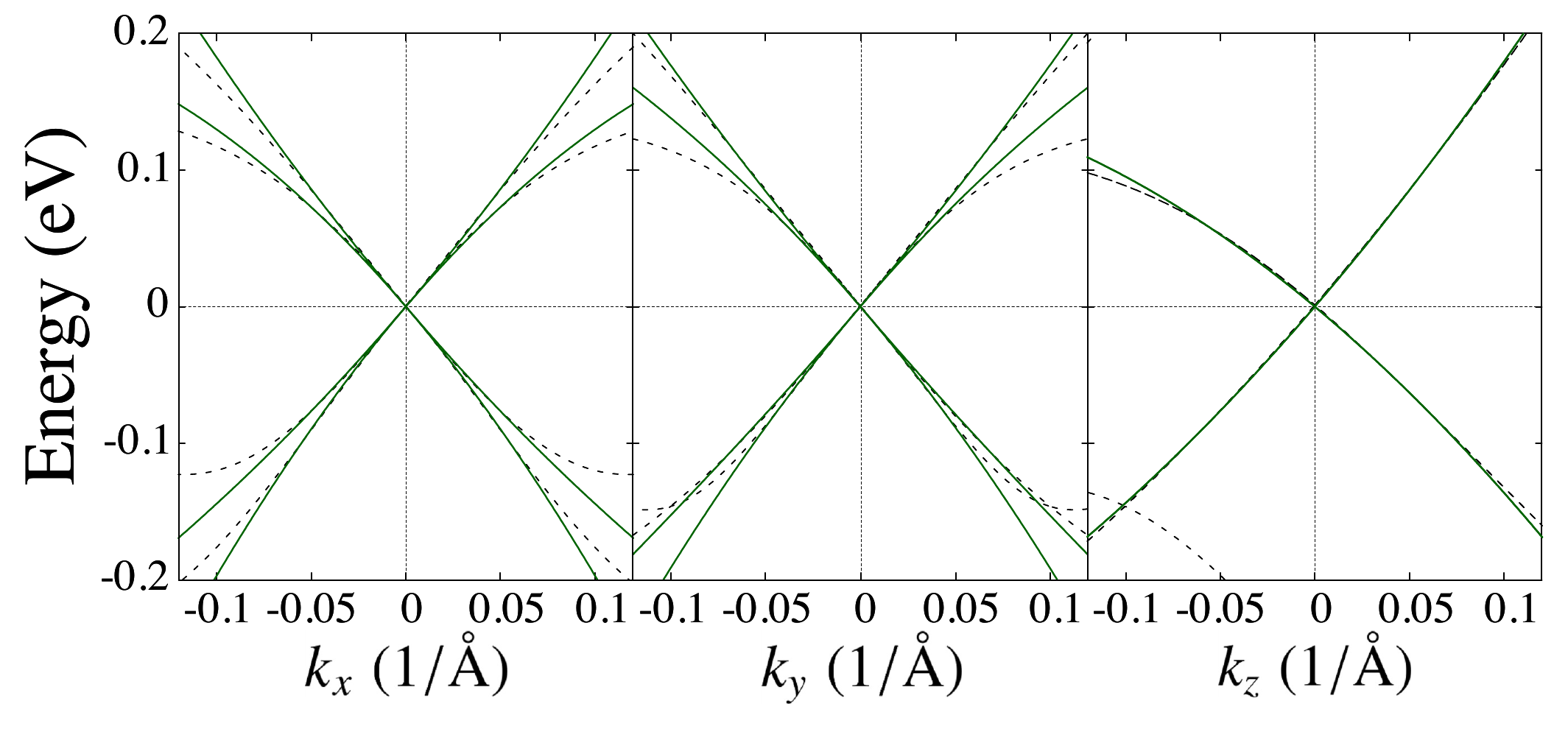}
\caption{\label{fig:kp_fitting}
\textbf{The comparison of the band structure of $\bm{\mathrm{Y}_8\mathrm{CoIn}_3}$ near the Dirac point between the $\bm{k \cdot p}$ model in Eq.~(2) of the main text and the first-principles calculations.}
The green solid line and the black dashed line represent the band structure of the $k\cdot p$ Hamiltonian with the parameters in Supplementary Table~\ref{tab:kp_fitting_parameter} and the first-principles calculations, respectively.
}
\end{suppfigure}

Supplementary Table~\ref{tab:kp_fitting_parameter} shows the fitted parameters of the Hamiltonian in Eq.~(2) of the main text.
These parameters reproduce well the band structure of the first-principles calculation in the vicinity of the Dirac point, as shown in Supplementary Fig.~\ref{fig:kp_fitting}.

\section*{Supplementary Note 5. Tight-binding model}
\label{sec:kpHamil_TB}

Here we present a tight-binding model on a hexagonal lattice with $C_{6v}$ symmetry and discuss some topological properties.

\begin{suppfigure}[htp]
\includegraphics[width=9cm]{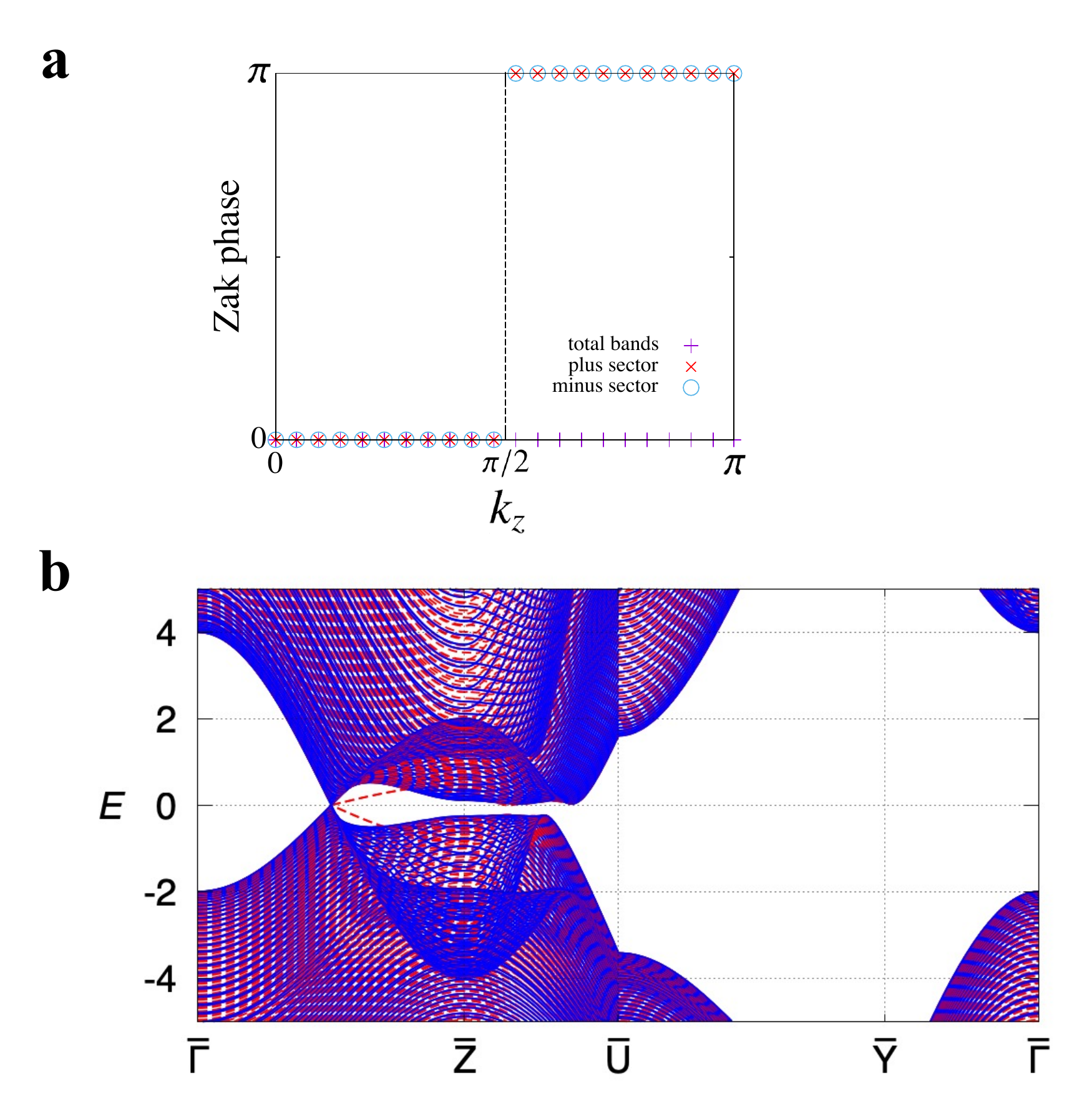}
\caption{\label{fig:kp_Zak_Surface}
\textbf{The Zak phases and surface states of the tight-binding Hamiltonian in Supplementary Eq.~(\ref{eq:kp_lat}), where the parameters are chosen as $\bm{(c_1, c_2, c_3, c_4, c_5, c_6, c_7, c_8, c_9, c_{10}, c_{11}, c_{12}) = (0, 0, 1, 0, 1, 1, 1, 0, 1, 3, 3, 1)}$.}
\textbf{a} The Zak phases along the $x$ axis for each glide sector.
The dashed line represents the coordinate of the Dirac point ($k_z = \pi/2$).
\textbf{b} The energy band structure calculated with a slab consisting of 128 unit cells along the $x$-axis.
The blue solid and red dashed lines represent the band structure calculated under the periodic and open boundary conditions, respectively.
}
\end{suppfigure}

\begin{suppfigure}[htp]
\includegraphics[width=9cm]{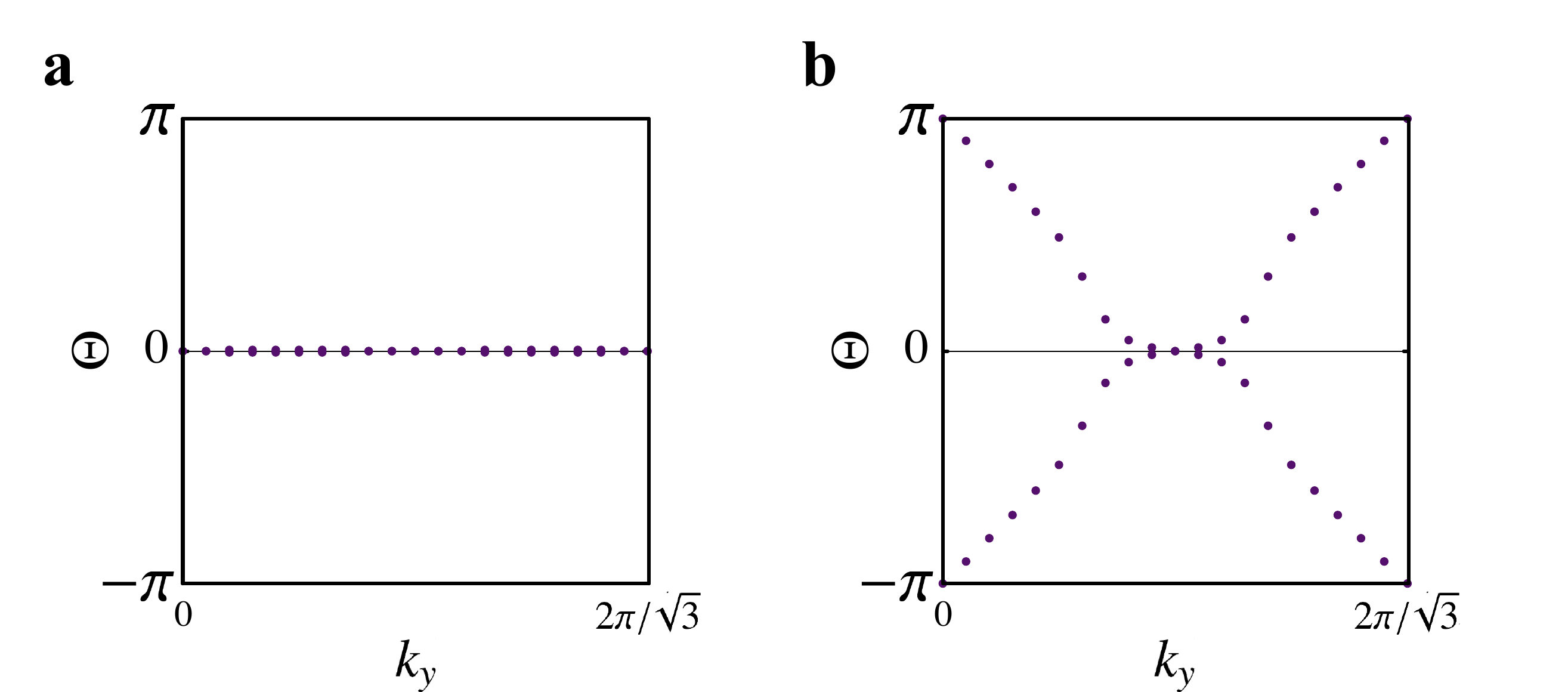}
\caption{\label{fig:kp_Wilson}
\textbf{The Wilson loop spectra of the tight-binding Hamiltonian in Supplementary Eq.~(\ref{eq:kp_lat}) for a $\bm{k_z = 0}$ and b $\bm{k_z = \pi}$ planes, where we choose the same parameters as in Supplementary Fig.~\ref{fig:kp_Zak_Surface}.}
The integral path of the Wilson loop is along the $k_x$ direction at fixed $k_y$.
}
\end{suppfigure}

The lattice Hamiltonian has the form
\begin{align}
\label{eq:kp_lat}
H_{\mathrm{lat}}(\bm{k}) &= c_1 \sin k_z \Gamma_{0,0} + c_2 \sin k_z \Gamma_{3,0} \nonumber\\
&\phantom{{}={}} + (c_3 + c_8 \sin k_z) [e_{11}(\bm{k}) \Gamma_{2,0} - e_{12}(\bm{k}) \Gamma_{1,2}] \nonumber\\
&\phantom{{}={}} + (c_4 + c_9 \sin k_z) [e_{11}(\bm{k}) \Gamma_{1,0} + e_{12}(\bm{k}) \Gamma_{2,2}] \nonumber\\
&\phantom{{}={}} + c_5 a_1(\bm{k}) \Gamma_{0,0} + c_6 \cos k_z \Gamma_{0,0} \nonumber\\
&\phantom{{}={}} + c_7 [e_{21}(\bm{k}) \Gamma_{3,3} - e_{22}(\bm{k}) \Gamma_{0,1}] \nonumber\\
&\phantom{{}={}} + c_{10} a_1(\bm{k}) \Gamma_{3,0} + c_{11} \cos k_z \Gamma_{3,0} \nonumber\\
&\phantom{{}={}} + c_{12} [e_{21}(\bm{k}) \Gamma_{0,3} - e_{22}(\bm{k}) \Gamma_{3,1}],
\end{align}
where
\begin{align}
    a_1(\bm{k}) &= 4 - \dfrac{4}{3} \left(\cos k_x + \cos\dfrac{-k_x+\sqrt{3}k_y}{2} \right. \nonumber\\
    &\phantom{{}={}} + \left. \cos\dfrac{-k_x-\sqrt{3}k_y}{2} \right), \\
    e_{11}(\bm{k}) &= \dfrac{2}{3} \qty(2\cos\dfrac{k_x}{2} + \cos\dfrac{\sqrt{3}k_y}{2}) \sin\dfrac{k_x}{2}, \\
    e_{12}(\bm{k}) &= \dfrac{2\sqrt{3}}{3} \cos\dfrac{k_x}{2} \sin\dfrac{\sqrt{3}k_y}{2}, \\
    e_{21}(\bm{k}) &= -\dfrac{8}{3} \qty(\cos k_x - \cos\dfrac{k_x}{2} \cos\dfrac{\sqrt{3}k_y}{2}), \\
    e_{22}(\bm{k}) &= \dfrac{8\sqrt{3}}{3} \sin\dfrac{k_x}{2} \sin\dfrac{\sqrt{3}k_y}{2}.
\end{align}
This Hamiltonian is identical to the $k \cdot p$ Hamiltonian in Eq.~(2) of the main text up to the second order in $k$.
Note that $\bm{k} = \bm{0}$ is the $\Gamma$ point in this lattice model, whereas it represents the position of the Dirac point in the $k \cdot p$ Hamiltonian.
Therefore, the tight-binding Hamiltonian in Supplementary Eq.~\eqref{eq:kp_lat} has to satisfy the time reversal symmetry $\Theta = \Gamma_{0,0} K$ ($K$: complex conjugation), which results in $c_1 = c_2 = c_4 = c_8 = 0$.
In these constraints, Dirac points are located at $(0,0,\pm \pi/2)$.

Supplementary Figure~\ref{fig:kp_Zak_Surface}a shows the Zak phase for each glide sector, where the integration is performed along the $k_x$ axis in the $k_y = 0$ plane.
As is the case in \YCI, the Zak phases in both sectors change by $\pi$ at the Dirac point.
To confirm the correspondence between the Zak phases and the surface states, we perform the Fourier transformation of the Hamiltonian in Supplementary Eq.~(\ref{eq:kp_lat}) with respect to the $k_x$ direction, and calculate the energy eigenvalues under the periodic and open boundary conditions.
As seen in Supplementary Fig.~\ref{fig:kp_Zak_Surface}b, in the open boundary condition, the two surface bands extend from the Dirac point towards the $\bar{Z}$ point, corresponding to the nontrivial Zak phases.

Since this model has the pure rotational symmetry rather than the screw, the $C_{2z}\Theta$-symmetry-protected topological number can be defined not only in the $k_z = 0$ plane but also in the $k_z = \pi$ plane.
The number of the occupied bands in this model is two, and thus the topological number is given by the $\mathbb{Z}$-valued Euler class~\cite{PhysRevB.99.235125,PhysRevB.102.115135,PhysRevB.104.195114}.
Supplementary Figure~\ref{fig:kp_Wilson} shows the Wilson loop spectra calculated in the $k_z = 0$ and $k_z = \pi$ planes, where the integral path of the Wilson loop is along the $k_x$ direction at fixed $k_y$.
As shown by Supplementary Fig.~\ref{fig:kp_Wilson}a, the Wilson loop spectrum in the $k_z = 0$ plane shows no windings, which indicates the trivial (zero) Euler class.
On the other hand, Supplementary Fig.~\ref{fig:kp_Wilson}b shows the Wilson loop spectrum with the winding number $+1$ and $-1$, which implies that the Euler class is 1, in the $k_z = \pi$ plane.
Therefore, the Dirac point in this system can be considered to emerge as a gap-closing object bridging the two planes with the different Euler classes.

\section*{Supplementary Note 6. $k\cdot p$ Hamiltonian with $C_{2v}$ symmetry}
\label{sec:kpHamil_C2v}

In this section, we discuss the nodal lines near $\bm{k} = \bm{0}$ analytically for the Hamiltonian in Eq.~(2) of the main text with the perturbation $c_{13}\Gamma_{0,3} + c_{14}\Gamma_{3,3}$, which reduces the symmetry to $C_{2v}$.
In the following, we assume that the perturbation is sufficiently small and that $\abs{c_{13}} \neq \abs{c_{14}}$.
In the following, we assume that the perturbation is sufficiently small and that $\abs{c_{13}} \neq \abs{c_{14}}$.

The eigenvalues on the $k_x = 0$ plane are
\begin{align}
&E(0, k_y, k_z) \nonumber\\
&= (-1)^{s_1} c_{14} + (c_5 - (-1)^{s_1} c_7) k_y^2 + (c_1 + c_6 k_z) k_z \nonumber\\
&\phantom{{}={}} + (-1)^{s_2} \left\{ (c_3 + c_8 k_z)^2 k_y^2 + (c_4 + c_9 k_z)^2 k_y^2 \right. \nonumber\\
&\phantom{{}={}} \left. + \left[(-1)^{s_1} c_{13} + c_2 k_z + c_{10} k_y^2 - (-1)^{s_1} c_{12} k_y^2 + c_{11} k_z^2\right]^2 \right\}^{1/2},
\end{align}
where $s_1, s_2 \in \{0, 1\}$.
When $c_{14} > 0$, the top valence and bottom conduction bands are degenerate at $\bm{k}$ that satisfy
\begin{align}
2 (c_{14} - c_7 k_y^2) &= \left[ (c_3 + c_8 k_z)^2 k_y^2 + (c_4 + c_9 k_z)^2 k_y^2 \right. \nonumber\\
&\phantom{{}={}} \left. + (c_{13} + c_2 k_z + c_{10} k_y^2 - c_{12} k_y^2 + c_{11} k_z^2)^2 \right]^{1/2} \nonumber\\
&\phantom{{}={}} + \left[ (c_3 + c_8 k_z)^2 k_y^2 + (c_4 + c_9 k_z)^2 k_y^2 \right. \nonumber\\
&\phantom{{}={}} \left. + (- c_{13} + c_2 k_z + c_{10} k_y^2 + c_{12} k_y^2 + c_{11} k_z^2)^2 \right]^{1/2}.
\end{align}
Eliminating the square roots by squaring both sides twice, we obtain
\begin{align}
\label{eq:C2v_vcNL}
&c_{14} \left\{c_{14} (c_3^2 + c_4^2) - 2 \left[(c_{13}^2 - 2 c_{14}^2) c_7 + c_{13} c_{14} c_{12}\right]\right\} k_y^2 \nonumber\\
&+ (c_{14}^2 - c_{13}^2) c_2^2 k_z^2 = c_{14}^2 (c_{14}^2 - c_{13}^2)
\end{align}
up to the second order in $k$.
To make this equation equivalent to the original one, we need 
\begin{align}
\label{eq:C2v_vcNL_c1}
c_{14} - c_7 k_y^2 > 0
\end{align}
and 
\begin{align}
\label{eq:C2v_vcNL_c2}
(c_3^2 + c_4^2 - 2 c_{12} c_{13} + 4 c_7 c_{14}) k_y^2 + c_2^2 k_z^2 < 2 c_{14}^2 - c_{13}^2.
\end{align}
No wavevector $\bm{k}$ satisfies Supplementary Eq.~(\ref{eq:C2v_vcNL_c2}) when $2 c_{14}^2 - c_{13}^2 < 0$, whereas Supplementary Eq.~(\ref{eq:C2v_vcNL_c2}) is satisfied inside the ellipse when $2 c_{14}^2 - c_{13}^2 > 0$.
If $\abs{c_{13}} > \abs{c_{14}}$, Supplementary Eq.~(\ref{eq:C2v_vcNL}) represents a hyperbola, and this exists outside the ellipse in Supplementary Eq.~(\ref{eq:C2v_vcNL_c2}).
Conversely, if $\abs{c_{13}} < \abs{c_{14}}$, Supplementary Eq.~(\ref{eq:C2v_vcNL}) represents an ellipse, and this lies inside the ellipse in Supplementary Eq.~(\ref{eq:C2v_vcNL_c2}).
Recalling that we assume $c_{14}> 0$, the ellipse in Supplementary Eq.~(\ref{eq:C2v_vcNL}) also satisfies Supplementary Eq.~(\ref{eq:C2v_vcNL_c1}) when $\abs{c_{13}} < \abs{c_{14}}$.
A similar result holds when $c_{14} < 0$.
Therefore, in the $k_x = 0$ plane, the top valence and bottom conduction bands are degenerate on the ellipse of Supplementary Eq.~(\ref{eq:C2v_vcNL}) when $\abs{c_{13}} < \abs{c_{14}}$.

On the other hand, the two valence bands are degenerate in the part of the curve
\begin{align}
\label{eq:C2v_vvNL}
&- c_{14} \left\{c_{14} (c_3^2 + c_4^2) -2 \left[(c_{13}^2 - 2c_{14}^2) c_7 + c_{13} c_{14} c_{12}\right]\right\} k_y^2 \nonumber\\
&+ (c_{13}^2 - c_{14}^2) c_2^2 k_z^2 = c_{14}^2 (c_{13}^2 - c_{14}^2)
\end{align}
that satisfies
\begin{align}
\label{eq:C2v_vvNL_c1}
c_{13} c_{14} (c_{10} k_y^2 + c_2 k_z) > 0
\end{align}
and
\begin{align}
\label{eq:C2v_vvNL_c2}
(c_3^2 + c_4^2 - 2 c_{12} c_{13} + 4 c_7 c_{14}) k_y^2 + c_2^2 k_z^2 > 2c_{14}^2 - c_{13}^2.
\end{align}
If $\abs{c_{13}} < \abs{c_{14}}$, Supplementary Eq.~(\ref{eq:C2v_vvNL}) describes an ellipse, and this does not fulfill Supplementary Eq.~(\ref{eq:C2v_vvNL_c2}).
On the contrary, if $\abs{c_{13}} > \abs{c_{14}}$, Supplementary Eq.~(\ref{eq:C2v_vvNL}) shows a hyperbola, and this satisfies Supplementary Eq.~(\ref{eq:C2v_vvNL_c2}).
The hyperbola in Supplementary Eq.~(\ref{eq:C2v_vvNL}) consists of two curves passing through $(0, 0, \pm c_{14}/c_2)$ respectively, but according to Supplementary Eq.~(\ref{eq:C2v_vvNL_c1}), only the one passing through $(0, 0, \sgn(c_{13}) c_{14}/c_2)$ becomes the nodal line.
A similar calculation for the two conduction bands shows that they are degenerate on the other curve of the hyperbola when $\abs{c_{13}} > \abs{c_{14}}$.
Note that the condition on the parameters $c_{13}$ and $c_{14}$ for the top valence and bottom conduction bands to be degenerate in the $k_x = 0$ plane is different from that for the valence bands and the conduction bands.

In the $k_y = 0$ plane, the top valence and bottom conduction bands are degenerate along the ellipse
\begin{align}
&c_{13} \left\{c_{13} (c_3^2 + c_4^2) + 2 \left[(c_{14}^2 - 2 c_{13}^2) c_{12} + c_{13}c_{14}c_7\right]\right\} k_x^2 \nonumber\\
&+ (c_{13}^2 - c_{14}^2) c_2^2 k_z^2 = c_{13}^2 (c_{13}^2 - c_{14}^2)
\end{align}
when $\abs{c_{13}} > \abs{c_{14}}$, while the two valence bands and the two conduction bands are degenerate along the hyperbola 
\begin{align}
&- c_{13} \left\{c_{13} (c_3^2 + c_4^2) + 2 \left[(c_{14}^2 - 2 c_{13}^2) c_{12} + c_{13} c_{14} c_7\right]\right\} k_x^2 \nonumber\\
&+ (c_{14}^2 - c_{13}^2) c_2^2 k_z^2 = c_{13}^2 (c_{14}^2 - c_{13}^2)
\end{align}
when $\abs{c_{13}} < \abs{c_{14}}$.

In summary, whether $\abs{c_{13}} < \abs{c_{14}}$ or $\abs{c_{13}} > \abs{c_{14}}$, the valence and conduction bands are degenerate along the ellipse, while the two valence bands and the two conduction bands are degenerate along the hyperbola, which lies on the plane orthogonal to that for the ellipse and passes through the interior of the ellipse.

\section*{Supplementary Note 7. $k\cdot p$ Hamiltonian with $C_{2}$ symmetry}
\label{sec:kpHamil_C2}

In this section, we add $c_{13}\Gamma_{3,1} + c_{14}\Gamma_{3,2}$ to the Hamiltonian in Eq.~(2) of the main text as an example of a perturbation that reduces the symmetry to $C_2$ , and discuss the Weyl points analytically.
The eigenstates on the $k_z$ axis are 
\begin{align}
\ket{\psi_1} &= \dfrac{1}{\sqrt{2}} \qty(\dfrac{-c_{13}+i c_{14}}{\sqrt{c_{13}^2 + c_{14}^2}}, 1, 0, 0)^\mathrm{T}, \\
\ket{\psi_2} &= \dfrac{1}{\sqrt{2}} \qty(0, 0, \dfrac{-c_{13}+i c_{14}}{\sqrt{c_{13}^2 + c_{14}^2}}, 1)^\mathrm{T}, \\
\ket{\psi_3} &= \dfrac{1}{\sqrt{2}} \qty(\dfrac{c_{13}-i c_{14}}{\sqrt{c_{13}^2 + c_{14}^2}}, 1, 0, 0)^\mathrm{T}, \\
\ket{\psi_4} &= \dfrac{1}{\sqrt{2}} \qty(0, 0, \dfrac{c_{13}-i c_{14}}{\sqrt{c_{13}^2 + c_{14}^2}}, 1)^\mathrm{T},
\end{align}
where $\mathrm{T}$ denotes the transpose.
When $c_2 > 0$, $\ket{\psi_1}$ and $\ket{\psi_2}$ are degenerate on the $k_z$ axis at 
\begin{equation}
k_z = k_{+} \coloneqq \dfrac{-c_2 + \sqrt{c_2^2 + 4 c_{11} \sqrt{c_{13}^2+c_{14}^2}}}{2 c_{11}} > 0, 
\end{equation}
and $\ket{\psi_3}$ and $\ket{\psi_4}$ are degenerate on the $k_z$ axis at 
\begin{equation}
k_z = k_{-} \coloneqq \dfrac{-c_2 + \sqrt{c_2^2 - 4 c_{11} \sqrt{c_{13}^2+c_{14}^2}}}{2 c_{11}} < 0.  
\end{equation}
The Hamiltonian projected onto the subspace spanned by $\ket{\psi_1}$ and $\ket{\psi_2}$ is
\begin{equation}
H_\mathrm{Weyl,+}(\bm{k}) = \sum_{i=0}^3 f_i(\bm{k})\sigma_i,
\end{equation}
where
\begin{align}
f_0(\bm{k}) &= (c_1 + c_6 k_z) k_z + \dfrac{2 c_7 c_{13} k_x k_y}{\sqrt{c_{13}^2 + c_{14}^2}} + c_5 (k_x^2 + k_y^2), \\
f_1(\bm{k}) &= (c_4 + c_9 k_z)k_x + \dfrac{c_{14} (c_3 + c_8 k_z)k_y}{\sqrt{c_{13}^2 + c_{14}^2}}, \\
f_2(\bm{k}) &= -(c_3 + c_8 k_z)k_x + \dfrac{c_{14} (c_4 + c_9 k_z)k_y}{\sqrt{c_{13}^2 + c_{14}^2}}, \\
f_3(\bm{k}) &= \sqrt{c_{13}^2 + c_{14}^2} - (c_2 + c_{11} k_z)k_z -\dfrac{2 c_{12} c_{13} k_x k_y}{\sqrt{c_{13}^2 + c_{14}^2}} \nonumber\\
&\phantom{{}={}} - c_{10} (k_x^2 + k_y^2).
\end{align}
The chirality of the Weyl point is calculated as
\begin{equation}
\sgn\qty( \det \left.\qty(\pdv{f_i}{k_j})_{i,j}\right|_{\bm{k}=(0,0,k_{+})} ) = -\sgn(c_{14}).
\end{equation}
By the same calculation, the chirality of the Weyl point created by $\ket{\psi_3}$ and $\ket{\psi_4}$ at $(0, 0, k_{-})$ is $\sgn(c_{14})$.
Thus, it is shown analytically that the addition of the perturbation $c_{13}\Gamma_{3,1} + c_{14}\Gamma_{3,2}$, which reduces the symmetry to $C_2$, produces the Weyl points of opposite chirality on the $k_z$ axis.
When $c_2 < 0$, the sign of the $k_z$ coordinate and the chirality of the Weyl points are reversed, but the above conclusion remains the same.
Note that the Weyl points are generated on the $k_z$ axis even when $c_{13} = 0$, in which case the system has $C_6$ symmetry.

\section*{Supplementary Note 8. Surface states in the various topological semimetal phases with the nontrivial second SW class}
\label{sec:Surface_states_SW}

\begin{suppfigure*}[htp]
\includegraphics[width=17cm]{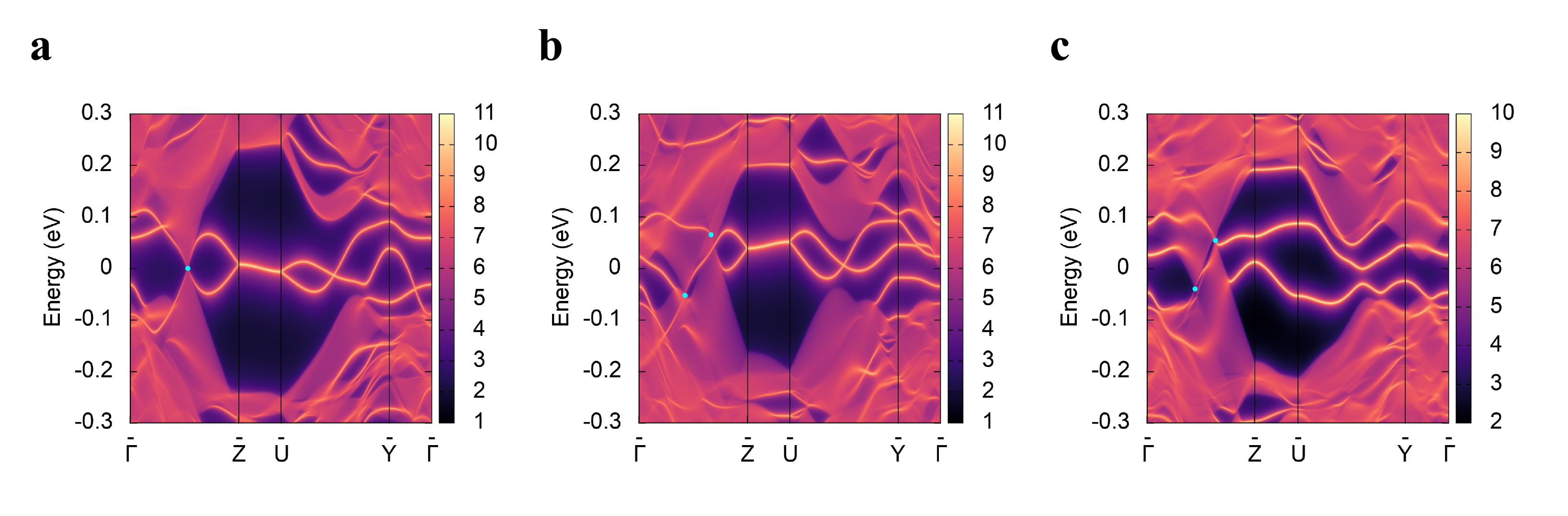}
\caption{\label{fig:Surface_transition_SW}
\textbf{Surface states of $\bm{\mathrm{Y}_8\mathrm{CoIn}_3}$ in the various topological semimetal phases characterized by the second SW class.}
The blue dots indicate points on the $k_z$ axis where the bulk valence and conduction bands are degenerate.
\textbf{a} Surface states for the (1000) surface of the original structure with the $C_{6v}$ symmetry. This is an enlarged version of Fig.~4b in the main text.
\textbf{b} Surface states of the distorted structure with the $C_{2v}$ symmetry.
\textbf{c} Surface states of the distorted structure with the $C_2$ symmetry.
}
\end{suppfigure*}

Here we discuss the surface states of \YCI in the various topological semimetallic phases that are characterized by the second SW class.

As shown in Supplementary Fig. \ref{fig:Surface_transition_SW}a, in the Dirac semimetal phase, the two midgap bands emerge from the projection of the bulk Dirac point. 
Their emergence corresponds to the Zak phases for each glide sector changing from $(0, 0)$ to $(\pi, \pi)$ at the Dirac point, as we discuss in the main text.

The compression of the lattice in the $x$ direction turns the system into the nodal-line semimetal phase, but the glide symmetry $\{M_y|00\frac{1}{2}\}$ is preserved. 
Recalling that the nodal line between the valence and conduction bands lies in the $k_x = 0$ plane, we can still discuss here the relationship between the Zak phases along the $x$ direction for each glide sector and the existence of nontrivial surface bands, as we did for the Dirac semimetal phase. 
In the nodal-line semimetal phase, the Zak phases for each glide sector are $(0, 0)$ near the $\Gamma$ point, $(0, \pi)$ inside the nodal line, and $(\pi, \pi)$ near the A point. 
Supplementary Figure \ref{fig:Surface_transition_SW}b shows that the number of topological surface states corresponds to these Zak phase values. 
Also, since the glide symmetry is maintained, the degeneracy of the midgap surface bands on the $\bar{Z}$-$\bar{U}$ line is preserved. 

As for the Weyl semimetal phase with $C_2$ symmetry, since there are Weyl points with opposite chirality on the $\Gamma$-A line, the surface bands emerge to connect their projections [Supplementary Fig. \ref{fig:Surface_transition_SW}c]. 
However, since the system no longer has mirror or glide symmetry, the Zak phases for each glide sector cannot be discussed, and the degeneracy on the $\bar{Z}$-$\bar{U}$ line is lifted.

\section*{Supplementary Note 9. Band structures of the magnetic material $\bm{\mathrm{Nd}_8\mathrm{CoGa}_3}$}

Here we present additional figures of the band structures of the magnetic material \ce{Nd8CoGa3}.

Supplementary Figures~\ref{fig:Nd8CoGa3_band_enlarge}a, b, c are enlarged versions of Figs.~7b, c, and d in the main text, respectively.
For comparison with the band structure in the DLM approach shown in Fig.~7d in the main text, we calculate the band structure using a pseudopotential that treats the Nd $4f$ orbitals as core states.
As shown in Supplementary Fig.~\ref{fig:Nd8CoGa3_band_enlarge}d, the result is in good agreement with the band structure obtained by the DLM method.
This agreement can be explained by the small spin splitting of itinerant Nd $4d$ electrons.

\begin{suppfigure*}[htp]
\includegraphics[width=16cm]{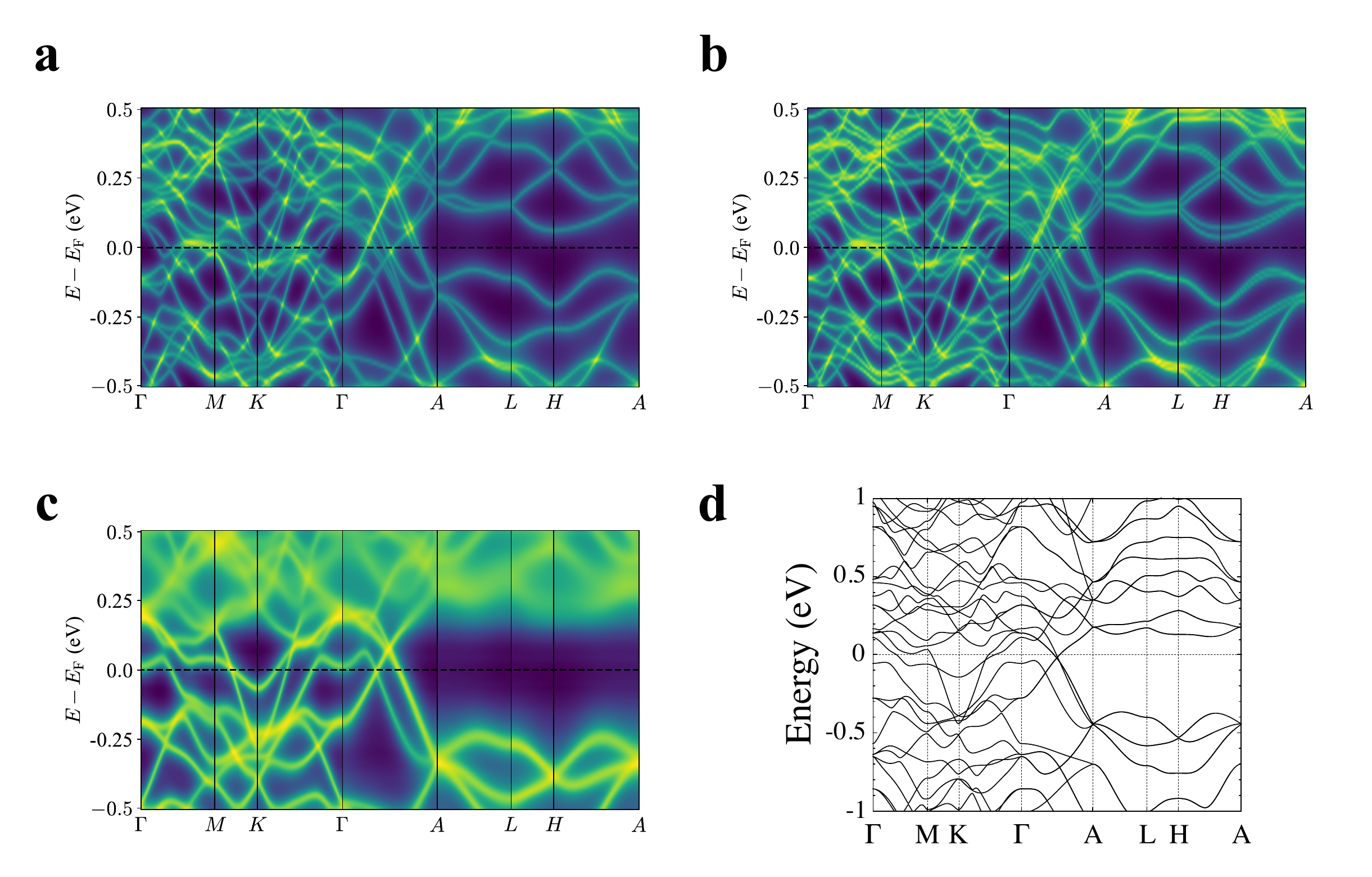}
\caption{\label{fig:Nd8CoGa3_band_enlarge}
\textbf{Band structures of the magnetic material \ce{Nd8CoGa3}.}
\textbf{a} The electronic band structure of \ce{Nd8CoGa3} without the SOC.
\textbf{b} The electronic band structure of \ce{Nd8CoGa3} with the SOC included self-consistently in ferromagnetic configurations.
\textbf{c} The electronic band structure of \ce{Nd8CoGa3} in the paramagnetic state computed using the DLM approach.
\textbf{a}, \textbf{b}, and \textbf{c} are enlarged views of Figs.~7b, c, and d in the main text, respectively.
\textbf{d} The electronic band structure of \ce{Nd8CoGa3} without the SOC calculated using a pseudopotential with the Nd $4f$ electrons in the core.
}
\end{suppfigure*}

We decompose the band structure in Fig.~7b in the main text with respect to the spin component. 
Supplementary Figures~\ref{fig:Nd8CoGa3_band_spin_decompose}a and b correspond to the majority and minority spin bands, respectively.

\begin{suppfigure*}[htp]
\includegraphics[width=16cm]{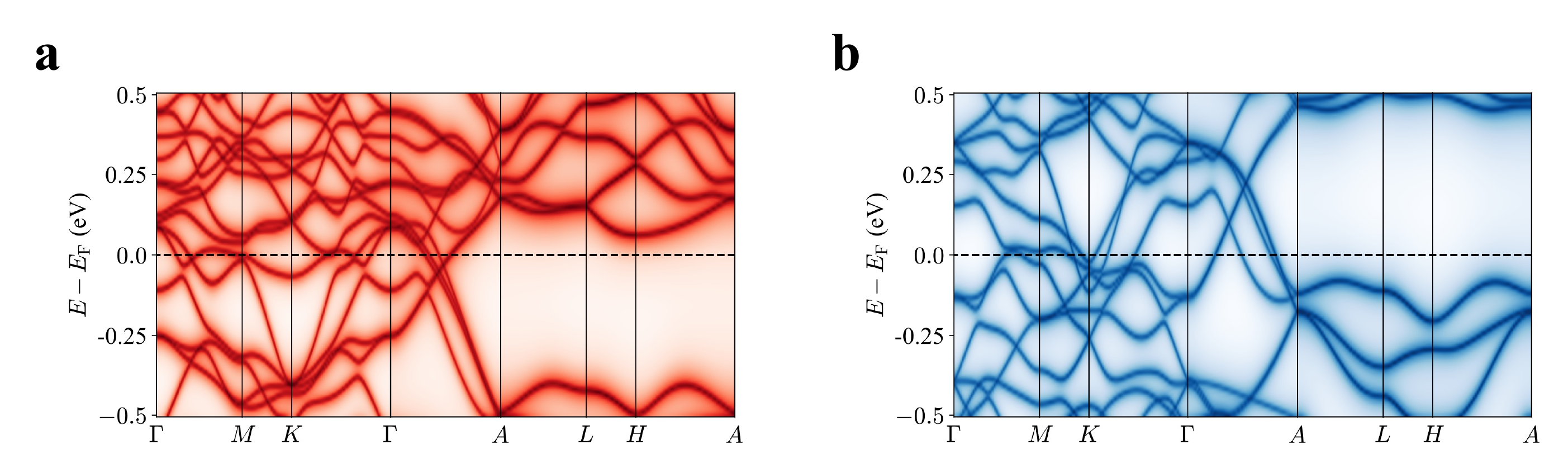}
\caption{\label{fig:Nd8CoGa3_band_spin_decompose}
\textbf{Band structures for each spin component of \ce{Nd8CoGa3}.}
These are obtained by decomposing Fig.~7b in the main text.
\textbf{a} Majority spin bands.
\textbf{b} Minority spin bands.
}
\end{suppfigure*}

\clearpage
%

\end{document}